\documentclass[twoside,leqnos]{article}  

\pdfoutput=0

\usepackage[margin=25mm]{geometry}

\usepackage{xspace}
\usepackage{paralist}
\usepackage{amsbsy,amssymb,amsmath,amsfonts}
\usepackage{fancybox}
\usepackage{epsfig}
\usepackage{color}
\usepackage{graphicx}
\usepackage{wrapfig}
\usepackage{caption}
\usepackage{subfig}
\usepackage{verbatim}

\usepackage{algorithm}
\usepackage{algorithmic}

\usepackage{pstricks}
\usepackage{pst-lens}

\usepackage{url}

\usepackage{ifpdf}

\DeclareGraphicsExtensions{.png}
\DeclareGraphicsExtensions{.jpg}
\newsubfloat[position=top, listofformat=subsimple]{algorithm}
\def\subalg{\subfloat}

\newcommand{\ignore}[1]{}

\newcommand{\runtime}{n\log n}

\newcommand{\Ms}{Minkowski~sum\xspace}
\newcommand{\Mswithbreak}{Minkowski sum\xspace}

\newcommand{\R}{\mathbb{R}}               
\newcommand{\Q}{\mathbb{Q}}               
\newcommand{\Z}{\mathbb{Z}}               

\newcommand{\cgal}{\textsc{Cgal}\xspace}

\newcommand{\eps}{\varepsilon}
\newcommand{\epsno}{\eps_{\mathrm{no}}}
\newcommand{\epsyes}{\eps_{\mathrm{yes}}}
\newcommand{\epsmid}{\eps_{\mathrm{mid}}}
\newcommand{\epshat}{{\hat{\eps}}}

\newcommand{\ddfrac}[2]{\frac{\displaystyle #1}{\displaystyle #2}}

\newcommand{\algdecidefull}{\textsc{Decide($Q,r,\eps$)}\xspace}
\newcommand{\algdecide}{\textsc{Decide}\xspace}
\newcommand{\algconvexdecidefull}{\textsc{ConvexDecide($Q,r,\eps$)}\xspace}
\newcommand{\algconvexdecide}{\textsc{ConvexDecide}\xspace}

\newcommand{\algdecideapp}{\textsc{ApproxDecide}\xspace}
\newcommand{\algdecideappinterior}{\textsc{ApproxDecideInterior}\xspace}
\newcommand{\algdecideappexterior}{\textsc{ApproxDecideExterior}\xspace}
\newcommand{\algdecideappfull}{\algdecideapp\textsc{($Q,r,\eps,\delta$)}\xspace}
\newcommand{\algdecideappinteriorfull}{\algdecideappinterior\textsc{($Q,r,\eps,\delta$)}\xspace}
\newcommand{\algdecideappexteriorfull}{\algdecideappexterior\textsc{($Q,r,\eps,\delta$)}\xspace}

\newcommand{\algsearchepsappfull}{\textsc{ApproxSearchEps($Q,r,\Delta$)}\xspace}

\newcommand{\annpol}[2]{\bar{D}_{{#1},{#2}}}
\newcommand{\criteps}{\hat{\eps}}

\newcommand{\critepsappd}{\widetilde{\eps_{\Delta}}}

\newcommand{\TP}{\Pi}
\newcommand{\tp}{u}

\newcommand{\boundary}{\partial}

\newcommand{\offset}{\mathrm{offset}}
\newcommand{\inset}{\mathrm{inset}}

\newcommand{\conv}{\mathrm{CH}}

\newcommand{\regterm}{eyelet}
\newcommand{\regstart}{spot}

\newcommand{\OPT}{\mathrm{OPT}}

\newcommand{\region}{\kappa}

\newcommand{\segment}[2]{\overline{{#1}{#2}}}

\newcommand{\YES}{{\small\sf YES\xspace}}
\newcommand{\NO}{{\small\sf NO\xspace}}
\newcommand{\UNDECIDED}{{\small\sf UNDECIDED\xspace}}

\usepackage{amsthm}

\newtheorem{definition}{Definition}
\newtheorem{lemma}[definition]{Lemma}
\newtheorem{theorem}[definition]{Theorem}
\newtheorem{corollary}[definition]{Corollary}
\newtheorem{proposition}[definition]{Proposition}

\newcommand{\myparagraph}[1]{\textbf{#1.}}

\newif\ifwithexperiments
\withexperimentsfalse

\numberwithin{equation}{section}
\numberwithin{figure}{section}

\usepackage{amssymb}
\usepackage{graphicx}

\urldef{\maileb}\path|eric@mpi-inf.mpg.de|
\urldef{\maildh}\path|danha@post.tau.ac.il|
\urldef{\mailmk}\path|mkerber@ist.ac.at|
\urldef{\mailrp}\path|rozapoga@post.tau.ac.il|
\urldef{\maildhrp}\path|{danha,rozapoga}@post.tau.ac.il|

\title{Deconstructing Approximate Offsets\footnote{A preliminary and partial version
appeared in  the Proceedings of the 27th annual ACM Symposium on
Computational Geometry (SoCG 2011)}.}

\author{
Eric Berberich\footnote{Max-Planck-Institut f\"ur Informatik, Saarbr\"ucken, Germany, \texttt{eric@mpi-inf.mpg.de}}\\
Dan Halperin\footnote{Tel Aviv University, Tel Aviv, Israel, \texttt{danha@post.tau.ac.il}}\\
\and
Michael Kerber\footnote{Institute of Science and Technology (IST) Austria, Klosterneuburg, Austria, \texttt{mkerber@ist.ac.at}}\\
Roza Pogalnikova\footnote{Tel Aviv University, Tel Aviv, Israel, \texttt{rozapoga@post.tau.ac.il}}\\
}
\date{}

\begin{document}
 
\maketitle

\begin{abstract}
We consider the \emph{offset-deconstruction problem}: Given
a polygonal shape~$Q$ with $n$ vertices, can it be
expressed, up to a tolerance~$\eps$ in Hausdorff distance,
as the Minkowski sum of another polygonal shape $P$ with a
disk of fixed radius? If it does, we also seek a preferably
simple-looking solution~$P$; then, $P$'s offset
constitutes an accurate, vertex-reduced, and smoothened
approximation of $Q$. We give an $O(\runtime)$-time exact
decision algorithm that handles any polygonal shape,
assuming the real-RAM model of computation. 
A variant of the algorithm, which we have implemented
using the \textsc{cgal} library, 
is based on rational arithmetic and answers the
same deconstruction problem up to an uncertainty parameter
$\delta$; its running time additionally depends on $\delta$.
If the input shape is found to be approximable,
this algorithm %
also computes an approximate solution for the problem.
It also allows us to solve parameter-optimization problems
induced by the offset-deconstruction problem.
For convex shapes, the complexity of the exact
decision algorithm
drops to $O(n)$, which is also the time required to compute
a solution~$P$ with at most one more vertex than a
vertex-minimal one.
\end{abstract}

\ignore{



\vspace{2mm}
\noindent
{\bf Categories and Subject Descriptors:} \\
{I.3.5} {[Computer Graphics]}: {Computational Geometry and Object Modeling} \\
{F.2.2} {[Analysis of Algorithms and Problem Complexity]}: {Nonnumerical Algorithms and Problems} - {\it Geometrical problems and computations}

\vspace{1mm}
\noindent
{\bf General Terms:} Algorithms, Theory

\vspace{1mm}
}

\section{Introduction}
\label{sec:introduction} The \emph{$r$-offset} of a
polygon, for a real parameter $r>0$, is the set of points
at distance at most $r$ away from the polygon. Computing
the offset of a polygon is a fundamental operation.  The
offset operation is, for instance, used to define a tolerance zone around
the given polygon~\cite{ha-nctool-1992} or
to dilute details for clarity of graphic 
exposition~\cite{matheron-morph-74,serra-83,dvks-building-generalization-2008}.
Technically, it is usually computed as the \emph{\Ms}of the
polygon and a disk of radius~$r$. The resulting shape is 
bounded by straight-line segments and circular arcs. 
However, a customary practice is to model the
disk in the \Ms{} with a (tight) polygon, which yields a
piecewise-linear approximation of the offset. 
Our study is motivated by two applications, where such an approximation
forms the legacy data which a program has to deal
with---the original shape before offsetting is unknown.
This leads us to the question what is the original polygon
whose approximate offset we have at hand. Of course,
finding the exact original polygon, or even its topology,
is impossible in general, because the offset might have
blurred small features like holes or dents. However, a
reasonable choice can lead to a more compact and smooth
representation of the approximate offset.

The first relevant problem concerns cutting polygonal parts out of
wood. A wood-cutting machine, which can smoothly cut along
straight line segments and circular arcs, is given a plan
to cut out a certain shape. This shape was designed as a
polygon expanded by a small offset, but with circular arcs
approximated by polygonal lines comprising many tiny line
segments. Thus instead of moving smoothly along circular
arcs, the cutting tool has to move along a sequence of very
short segments, and make a small turn between every pair of
segments. The process becomes very slow, the tool heats up,
and occasionally it causes the wood to burn.
Moving the cutting tool smoothly and fast enough is the way
to keep it cool. If this were the only issue, other
smoothing techniques like \emph{arc-spline approximation}
\cite{Drysdale200831,HeimlichH08} may have been applicable.
However, we may also wish to reduce the offset radius if a
more accurate cutting machine is available---in this case,
it seems desirable to find the original shape first and
then to re-offset with a smaller radius.

A motivation to study this question from a different domain
is to recover shapes sketched by a user of a digital pen
and tablet. The pen has a relatively wide tip, and the input
obtained is in fact an approximate offset (with the radius
of the pen tip) of the intended shape. The goal is to give
a good polygonal approximation of the intended shape.
%
More broadly, as the offset operation is so
commonplace, it seems natural to ask, given only an
(approximated) offset shape, what could be the original
shape before the offsetting.
Therefore, we pose the \emph{(offset-)deconstruction problem}
which comes in two variants:

\begin{compactitem}
\item [\textbf{Problem~1: the decision problem}]~\\ Given a polygonal shape~$Q$, and two
real parameters $r,\eps > 0$, decide if there exists a polygonal shape $P$
such that $Q$ is within (symmetric) 
Hausdorff-distance $\eps$ to the $r$-offset (i.e., offset with radius $r$)
of $P$
\item [\textbf{Problem~2: finding a solution}]~\\ If the answer to Problem~1 is \YES, compute a
polygonal shape $P$ with the desired property. We refer to $P$ as a \emph{solution} of the deconstruction problem. Note that $P$ might be disconnected, even if $Q$ is connected (Figure~\ref{fig:bsops}).
\end{compactitem}

Problem~1 can be seen as a special case of the
\emph{Minkowski decomposition problem} which asks whether a
set can be composed in a non-trivial way as the Minkowski
sum of two sets---disallowing a summand to be a homothetic
copy of the input set. A general criterion for
decomposability of convex sets in arbitrary dimension has
been presented in~\cite{sallee1972}. A particularly
well-studied case are planar lattice polygons, because of
their close relation to problems in algebra, for instance,
polynomial factorization~\cite{ostrowski-bedeutung}. It has
been shown that deciding decomposability is NP-complete for
lattice polygons~\cite{gl-decomposition}.
In~\cite{et-mindecomp-2006}, decomposability is
investigated under the constraint that one of the summands
is a line segment, a triangle, or a quadrangle. However,
all these approaches discuss the exact decomposition
problem; our scenario of being Hausdorff-close to a
particular decomposition seems to not have been addressed
in the literature.
Allowing tolerance raises interesting algorithmic questions
and at the same time makes the tools that we develop more
readily suitable for applications, which typically have to
deal with inaccuracies in measuring and modeling.

\newif\ifwithfigpics
\withfigpicstrue

\ifwithfigpics

\begin{figure*}[!htp]
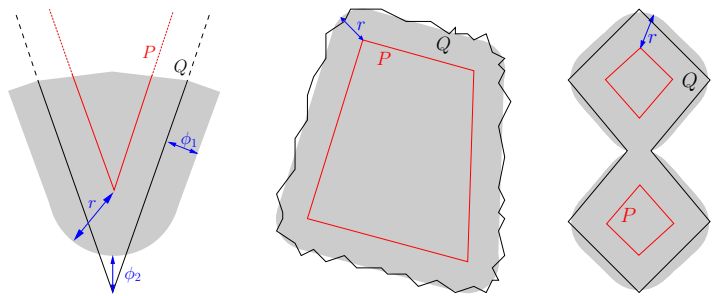

  \centering
\hspace{-3mm}\resizebox{!}{3.8cm}{\input{spikes_are_bad.pstex_t}}
\hspace{0.5cm}
\resizebox{!}{3.8cm}{\input{Smoother.pstex_t}}
\hspace{0.5cm}
\resizebox{!}{3.8cm}{\input{P_in_two_comp.pstex_t}}
  \caption{For a given $Q$, the red $P$ is a candidate summand whose exact
$r$-offset is shaded.
Left: For a given $\eps$,
deconstruction is ensured iff $\phi_1 \leq \eps$ and $\phi_2 \leq \eps$.
Note that, when $r$ decreases, $\phi_1$ decreases, but $\phi_2$ increases.
Middle: Example where $Q$ can be approximated by an $r$-offset of a~$P$ that has
much fewer vertices than~$Q$.
Right: Example where
$Q$ can be approximated by the $r$-offset of a disconnected shape~$P$.}
\vspace{-0.4cm}
  \label{fig:bsops}
\end{figure*}

\else

\begin{figure}[!htp]
  \centering
\subfloat[][Input 1 - yes]{
\includegraphics[bb=0 0 187 187, width=2.5cm]{Figs/workc0.png}
\includegraphics[bb=0 0 187 187, width=2.5cm]{Figs/workc2.png}
}\,
\subfloat[][Input 2 - no]{
\includegraphics[bb=0 0 187 187, width=2.5cm]{Figs/failc0.png}
\includegraphics[bb=0 0 187 187, width=2.5cm]{Figs/failc2.png}
}\\
  \caption{Are the given shapes (blue) approximate $1$-offsets of
other polygonal shapes up to Hausdorff-distance~$\frac{1}{10}$? The black box
in the interior is the exact $\frac{9}{10}$-inset of the shapes - a possible
summand - at least for Input~1, as its approximated circular arcs
are close enough to the black box,
while the approximation of Input~2 does no fulfill this criterion.}
  \label{fig:bsops}
\end{figure}

\fi

\myparagraph{Contributions}
We first present an efficient algorithm to decide Problem~1:
For a shape $Q$ with $n$ vertices,
the algorithm reports the correct answer in
$O(\runtime)$ time in the real-RAM model of computation~%
\cite{ps-cgi-90}. It constructs offsets with increasing
radii in three stages; the intermediate shapes arising
during the computation are in general more difficult to
offset than polygons, as they are bounded by straight-line
segments and ``indented'' circular arcs (namely, the shape
is locally on the concave side of the arcs). The main
observation is that for certain classes of such shapes, these
circular arcs can be ignored when computing the next offset
(see Theorem~\ref{thm:concave-offset} for the precise
statement). This observation bounds the time required by
each offset computation by $O(\runtime)$, which is the key
to the efficiency of the decision algorithm. 
Our proof is constructive, that is, if a solution exists it can be computed
with the same running time.

The computation of the exact decision procedure requires
the handling of algebraic coordinates of considerably high
degree. As an alternative, we give an approximation scheme
that  works exclusively with rational numbers. The
scheme proceeds by replacing the offset disks by polygonal
shapes of similar diameter, whose precision is determined
by another parameter $\delta<\eps$. We prove a bound $\Delta$
that depends on $\epshat$, the minimal $\eps$ for which the
answer to the decision problem is \YES, such that the
rational approximation returns the exact result for all
$\delta\leq \Delta$. If the input shape is found to be
deconstructible, this algorithm also outputs a solution.
The computation of $\epshat$ up to any desired precision
is still possible. We believe that our investigation of the 
relation between $\delta$ and $\epshat$ is of independent 
relevance, mostly to the study of certified algorithms 
that approximate geometric objects with algebraic coordinates 
by means of rational arithmetic.

The deconstruction problem leads to natural optimization questions:
if $Q$ and $r$ are given, how to compute $\epshat$, the minimal
tolerance for which a solution exists? Similarly, 
if $Q$ and $\eps$ are given, what are the possible radii such that
a solution exists? For the first question, we provide
a certified and efficient solution based on binary search, using 
the rational approximation algorithm.
For the second question, we prove that the set of possible radii
forms an interval and propose an algorithm to compute it. 
We also provide a heuristic to find a reasonable
radius $r$ if both $r$ and $\eps$ are unknown.

For a convex shape $Q$ with $n$ vertices, we reduce the
running time for solving Problem~1 to the optimal $O(n)$
(in the real-RAM model). Moreover, we describe a greedy
algorithm within the same time complexity that 
returns a solution $P^\star$ which minimizes, up to
one extra vertex, the number of vertices among all
solutions, if there are any. Our algorithm technically
resembles an approach for the different problem of finding
a vertex-minimal polygon in the annulus of two nested
polygons~\cite{abrsy-fmcnp-89}. We also remark that the
$r$-offset of $P^\star$ has a tangent-continuous boundary
and therefore constitutes a special case of an arc-spline
approximation of $Q$ where all circular arcs have the same
radius.

\myparagraph{Organization}
We describe an exact decision algorithm for the
deconstruction problem (solving Problem~1 above) 
in Section~\ref{sec:decision}.
In Section~\ref{sec:rational_approx} we describe a
rational-approximation algorithm for the deconstruction
problem. 
Both algorithms output a solution in case the input is 
deconstructible (solving Problem~2).
Section~\ref{sec:search} discusses the optimization problems.
For convex input, Section~\ref{sec:convex} exposes a
specialized deconstruction algorithm and the computation of 
an almost vertex-minimal solution. 
We conclude in Section~\ref{sec:conclusion} by pointing 
out open problems.

\section{The Decision Algorithm}
\label{sec:decision}

For a set $X\subset\R^2$ denote its boundary by~$\boundary
X$ and its complement by~$X^C:=\R^2\setminus X$. 
A \emph{polygonal region} or \emph{polygonal shape}
$X\subset\R^2$ is a set whose boundary 
consists of finitely many line segments with disjoint interiors. 
The endpoints of these
straight-line segments are the \emph{vertices} of
the polygonal region. We assume henceforth that the input
shapes that we deal with are bounded (but not necessarily connected). 
Although the
techniques seem to go through also for unbounded shapes,
this assumption simplifies the exposition and is sufficient
for the real-life applications we have in mind. 
For two sets $X$ and $Y$, we denote their \emph{Minkowski
sum} by $X\oplus Y:=\{x+y\mid x\in X, y\in Y\}$. 
With $d(\cdot,\cdot)$ the
Euclidean distance function,
and any $c\in \R^2, r\in\R$, we write $D_r(c):=\{p\in\R^2\mid
d(c,p)\leq r\}$ for the (closed) $r$-disk around $c$, and
$D_r:=D_r(O)$ for the $r$-disk centered at the origin. The
\emph{$r$-offset} of a set $X$, $\offset(X,r)$, is the \Ms\
$X\oplus D_r$.

For $p\in\R^2$ and $X$ a closed set, 
we write $d(p,X):=\min\{d(p,x)\mid x\in X\}$.
The \emph{(symmetric) Hausdorff distance} of two closed point sets $X$ and $Y$
is 
$H(X,Y):=\max\{\max\{d(x,Y)\mid x\in X\}, \max\{d(y,X)\mid y\in Y\}\}.$
\emph{We say that $X$ is $\eps$-close to $Y$ (and $Y$ to $X$) if
$H(X,Y)\leq\eps$}, which can also be expressed
alternatively:
\begin{proposition}\label{prop:hausdorff}
For $X,Y$ closed, $X$ is $\eps$-close to~$Y$  if and only if
$Y\subseteq \offset(X,\eps)$ and $X\subseteq \offset(Y,\eps)$.
\end{proposition}

\myparagraph{Decision algorithm} We fix $r>0$, $\eps>0$,
and a polygonal region~$Q$, and consider the following
question: Is there a polygonal region $P$ such that $Q$
and the $r$-offset of $P$ have Hausdorff-distance at most
$\eps$? First of all, we can assume that $r>\eps$;
otherwise, we can choose $P:=Q$, because $\offset(Q,r)$
and~$Q$ have Hausdorff-distance at most~$\eps$. We define
another operation, \emph{$r$-inset} (a.k.a.~``erosion''),
which is computationally similar to an offset:
\begin{definition}
For $r>0$, and $X\subset\R^2$, the \emph{$r$-inset}
of $X$ is the set
$\inset(X,r):=\offset(X^C,r)^C=
\left\{ x\in\R^2\mid D_r(x)\subseteq X\right\}.$
\end{definition}

We are now ready to present the decision algorithm: 

\begin{algorithm}[H]
\caption{\textsc{\algdecidefull}}
\begin{compactenum}[(1)]
\item $Q_\eps\gets\offset(Q,\eps)$
\item $\TP\gets\inset(Q_\eps,r)$
\item $Q'\gets\offset(\TP,r+\eps)$
\item \textbf{if} $Q\subseteq Q'$ \textbf{then} return \YES\ \textbf{else} return \NO
\end{compactenum}
\label{alg:decision_general}
\end{algorithm}

%
We next prove that
\algdecide~(Algorithm~\ref{alg:decision_general}) correctly
decides whether $Q$ is $\eps$-close to some $r$-offset of a
polygonal region. A first observation is that for any
polygonal region~$P$, $\offset(P,r)\subseteq Q_\eps$ if and
only if $P\subseteq\TP$. This is an immediate consequence
of the definition of the inset operation. This shows that
for any $\offset(P,r)$ that is $\eps$-close to~$Q$, $P$
must be inside $\TP$. Moreover, it shows that any choice of
$P\subseteq \TP$ already satisfies one of
Proposition~\ref{prop:hausdorff}'s inclusions. It is only
left to check whether $Q\subseteq
\offset(\offset(P,r),\eps)=\offset(P,r+\eps)$. We
summarize:

\begin{proposition}\label{prop:closeness}
$Q$ is $\eps$-close to $\offset(P,r)$ if and only if
$P\subseteq\TP$ and $Q\subseteq\offset(P,r+\eps)$.
\end{proposition}

To prove correctness of \algdecide, we have to show that
$Q\subseteq\offset(\TP,r+\eps)$ already implies that there also
exists a polygonal region $P\subseteq\TP$ with
$Q\subseteq\offset(P,r+\eps)$. The main difficulty in proving this is 
that $\TP$ is not polygonal in general; we have to study its shape closer to
prove that we can approximate it by a polygonal region,
maintaining the property that the offset remains
$\eps$-close to~$Q$.

\myparagraph{The shape of offsets and insets}
For a polygonal region $Q$, it is not hard to figure out
the shape of $Q_\eps=\offset(Q,\eps)$: It is a closed
set bounded by straight-line segments and by circular
arcs, belonging to a circle of radius $\eps$. It is
important to remark that all circular arcs are
\emph{bulges}:

\begin{definition}
Let $X\subset\R^2$ be a closed set with some circular arc $\gamma$
on its boundary. Then, $\gamma$ is called a \emph{dent} with respect to
$X$, if each line segment connecting two distinct points 
on $\gamma$ is not fully
contained in $X$. Otherwise, the arc is called a \emph{bulge}.

We call $X$ a \emph{bulged} (resp.\ an \emph{indented})
\emph{region with radius $r$}, if $\boundary X$ consists of finitely many
straight-line segments and bulges (resp.\ dents) that are all of radius~$r$,
interlinked at the \emph{vertices} of the region.
\end{definition}

\begin{wrapfigure}[]{r}{3.0cm}
\includegraphics[width=3.0cm]{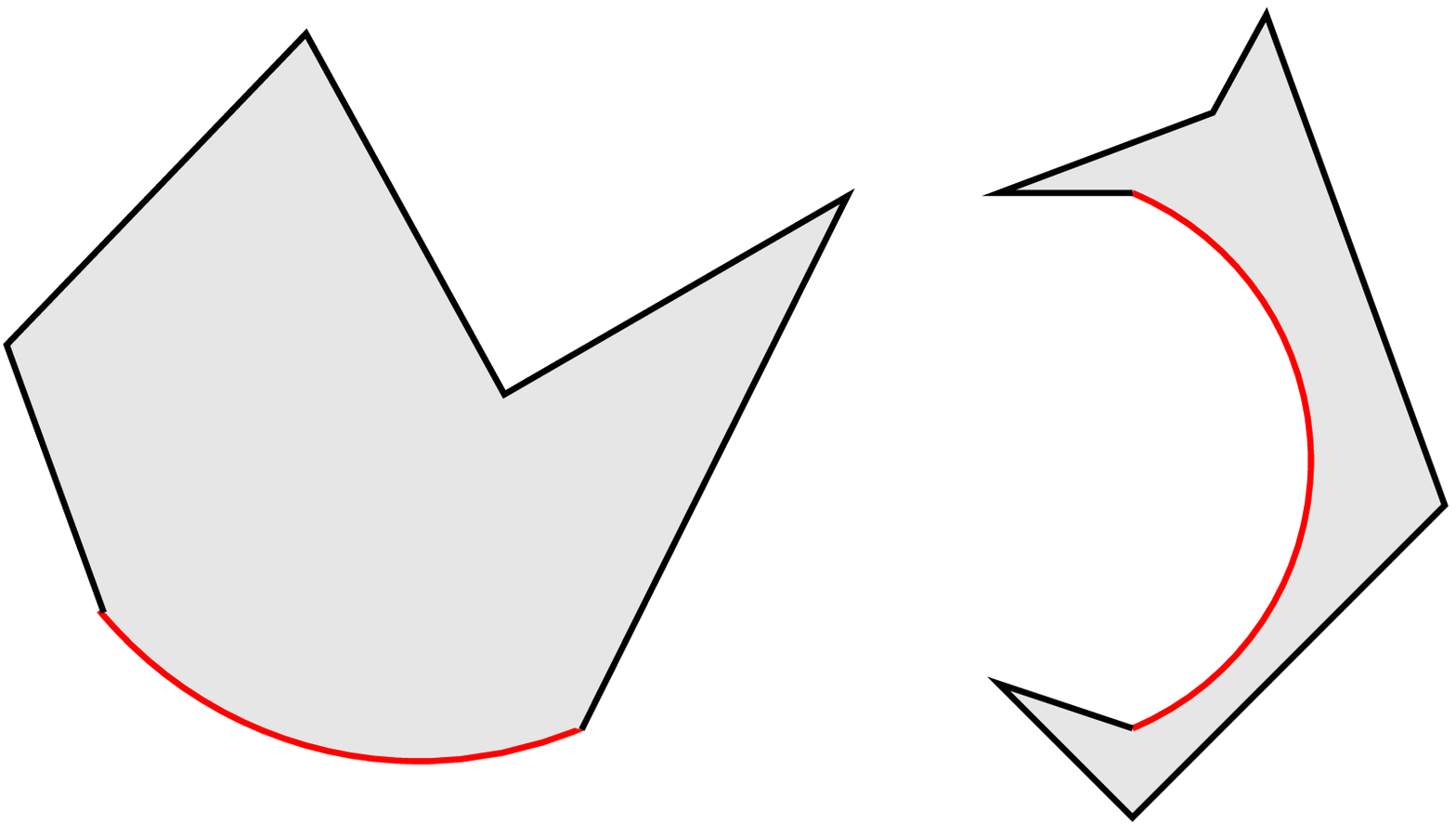}
\end{wrapfigure}
Note that a bulged region (left) is not necessarily
convex. The $r$-offset of a polygonal region $P$ is a
bulged region with radius~$r$. The heart of this
section is Theorem~\ref{thm:concave-offset} showing that the
same also holds if $P$ is an indented region (right)
with radius smaller than $r$:

\begin{theorem}\label{thm:concave-offset}
Let $P$ be an indented region with radius $r_1$,
and let $r_2>r_1$. Then, there is a polygonal region
$P_L\subseteq P$ such that
$\offset(P,r_2)=\offset(P_L,r_2)$. In particular,
$\offset(P,r_2)$ is a bulged region with radius~$r_2$.
\end{theorem}

\begin{figure*}[ht]
  \centering
\ignore{
\subfloat[]{\label{fig:concave_casea}\includegraphics[width=0.24\textwidth]
{linear_cap.eps}}
\subfloat[]{\label{fig:concave_caseb}\resizebox{0.24\textwidth}{!}{\input{
concave_case_1.pstex_t}}}
\subfloat[]{\label{fig:concave_cased}\resizebox{0.24\textwidth}{!}{\input{
concave_case_3.pstex_t}}}
}
\subfloat[]{\label{fig:concave_case1}\includegraphics[width=0.25\textwidth]
{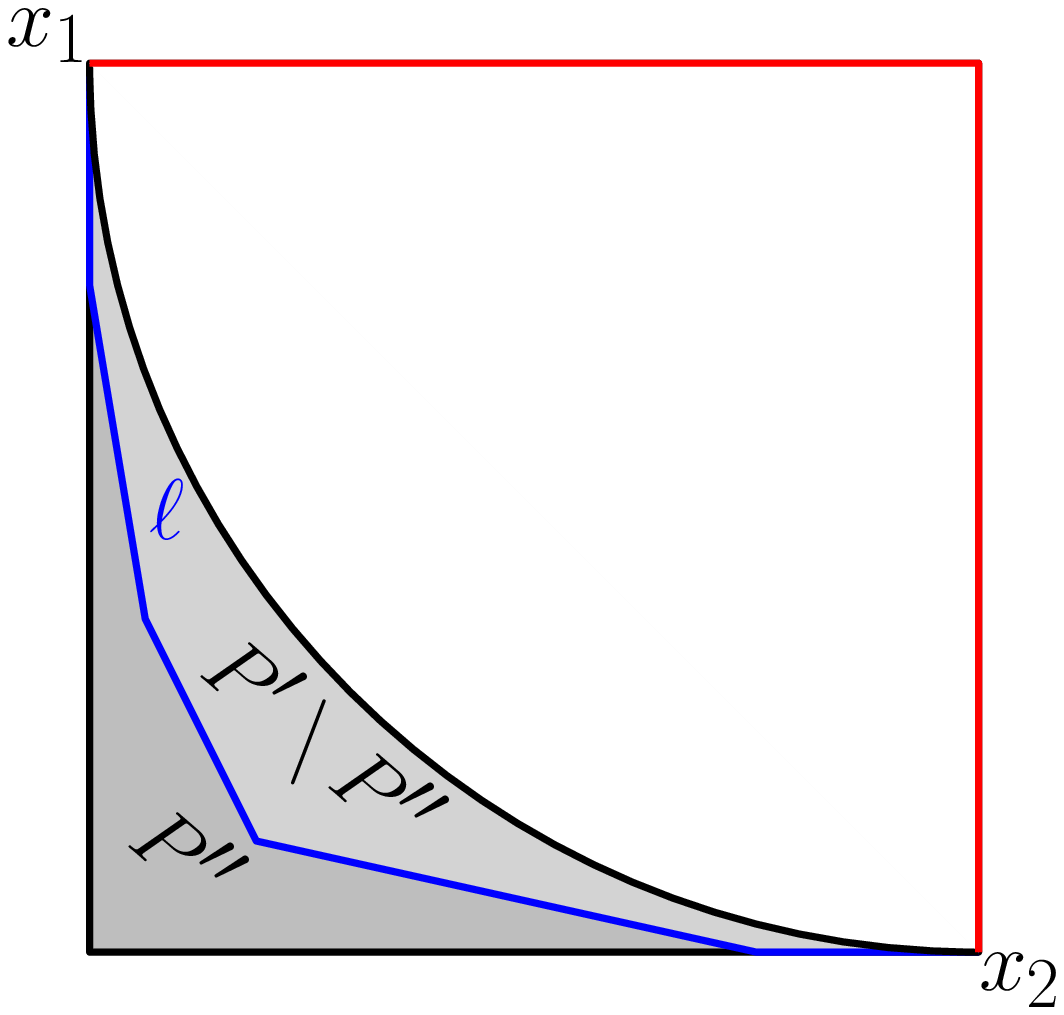}}
\subfloat[]{\label{fig:concave_case2}\includegraphics[width=0.25\textwidth]
{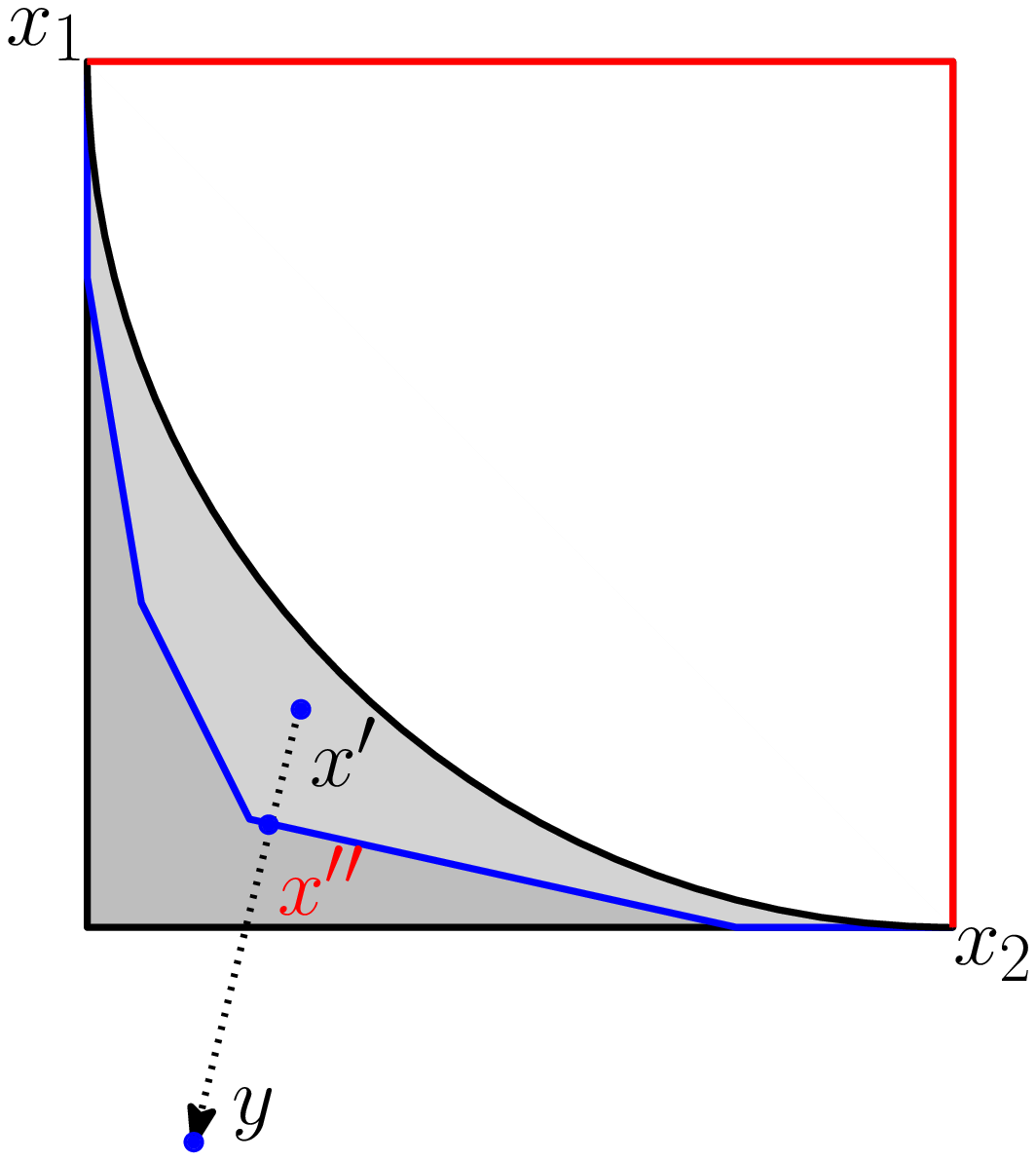}}
\subfloat[]{\label{fig:concave_case3}\includegraphics[width=0.25\textwidth]
{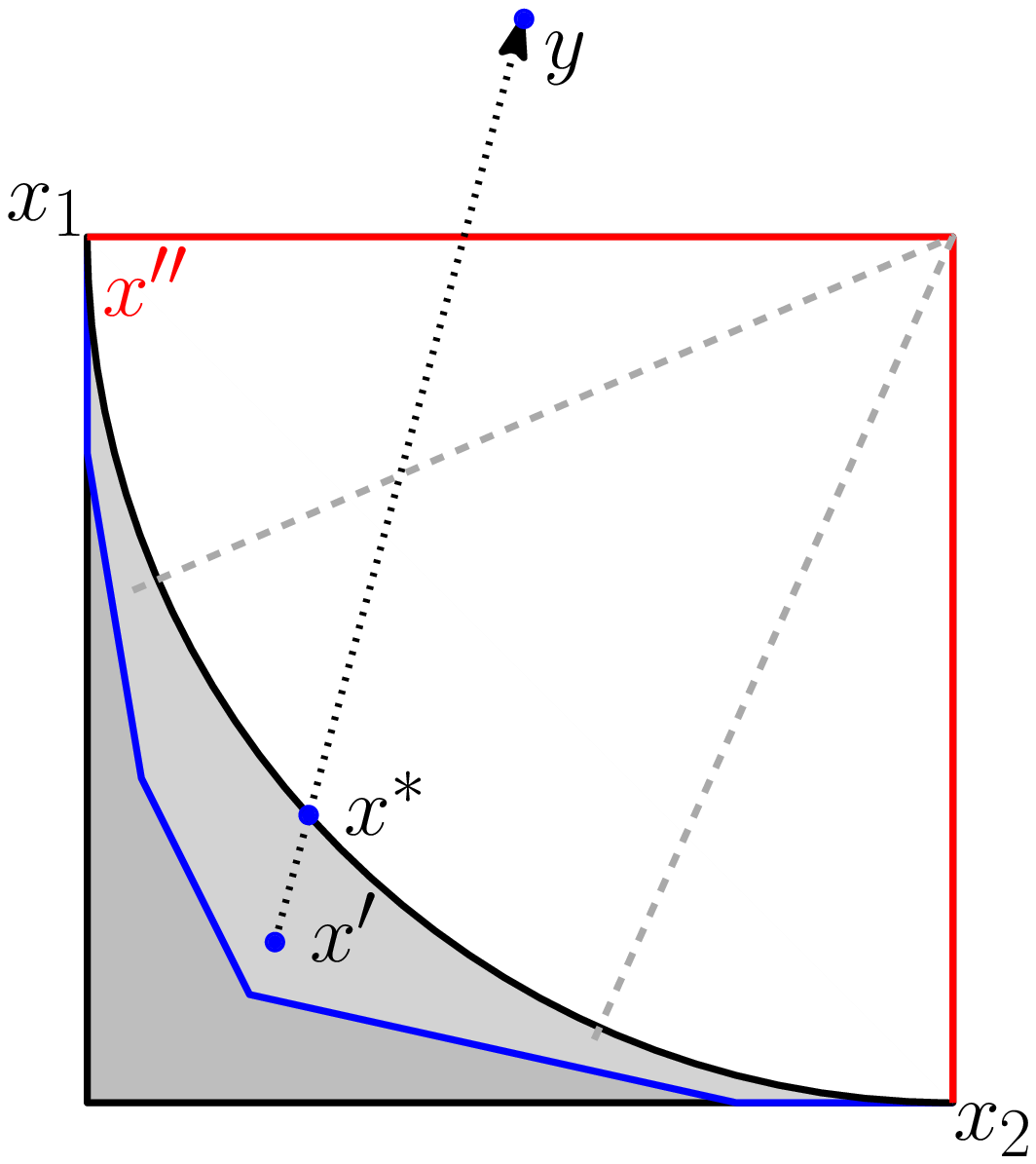}}
  \caption{ (a) The (extended) linear cap split by the polyline $\ell$, (b-c) the two cases in the
proof of Theorem~\ref{thm:concave-offset}.}
  \label{fig:concave_case}
\vspace{-0.5cm}
\end{figure*}

\begin{proof}
After possibly splitting circular arcs into at most four 
parts, we can assume that each circular arc spans at most 
a quarter of the circle. For such a circular arc, we define 
its endpoints by $x_1$ and $x_2$,
and denote the \emph{linear cap} of the
circular arc as the (closed) indented region enclosed by the
circular arc, and the two lines tangent to the circle
through $x_1$ and $x_2$ (the shaded area in Figure~\ref{fig:concave_case1}). 
The \emph{extended linear cap} is the (polygonal) region 
spanned by the two tangents just mentioned, and the two corresponding 
normals at $x_1$ and $x_2$. Clearly, the
normals meet in the center of the circle that defines the
arc.

We iteratively replace a indented arc of an indented region
$P'$ with radius $r_1$ (initially set to $P$) 
by a polyline $\ell$ ending in the endpoints of the circular
arc, such that $\ell$ does neither leave $P'$ nor the
linear cap of the circular arc, and such that other
boundary parts of $P'$ are not intersected. This yields
another indented region $P''$ with radius $r_1$, where one indented arc is
replaced by a polyline, as depicted in Figure~\ref{fig:concave_case1}. 
Iterating this construction,
starting with~$P$, until all indented arcs are replaced, we
obtain a polygonal region~$P_L$.

We show that in each iteration, the $r_2$-offsets of $P'$
and $P''$ are the same. For that we consider any point
$x'\in P'\setminus P''$, in the region that is cut off by
$P''$, and consider $y=x'+v'$ for an arbitrary $v'\in
D_{r_2}$. We show that in all cases, $y$ can also be
written by $y=x''+v''$, with $x''\in P''$, and $v''\in
D_{r_2}$. 

Since the circular arc spans at most a quarter of the circle, 
it is easily seen that $D_{r_1}(x_1)\cup D_{r_1}(x_2)$ 
covers the whole extended linear cap. Therefore,
for any $y$ that lies within the extended linear cap,
selecting $x''=x_1$ or $x''=x_2$, we get
$y=x''+v''$ with $v''\in D_{r_1}$.

We distinguish two other cases: for $y$ that lies 
outside of the extended linear cap $v'=\overline{x'y}$ crosses either $\ell$
or the circular arc. In the former case, we can simply 
pick the crossing point as $x''$, and set $v''\in D_{r_2}$ accordingly
(Figure~\ref{fig:concave_case2}). 
In the latter case, let us denote the crossing point as $x^*$
(Figure~\ref{fig:concave_case3}).
We consider the set of points that is closer to $x^*$ than to $x_1$ and $x_2$.
Clearly, that region is bounded by the two corresponding bisectors, which
meet in the center of the circle that defines the circular arc and is 
therefore completely contained within the extended linear cap.
It follows that $y$ is closer to one of the endpoints of the arc, say $x_1$,
than to $x^*$. Selecting $x''=x_1$ we ensure $y$ is closer to $x''$ than to $x'$, 
which proves that $y=x''+v''$ with some $v''\in D_{r_2}$ in this case as well.
\end{proof}

The proof of Theorem~\ref{thm:concave-offset} implies
that $\offset(P,r_2)$ for such a region $P$
is completely determined by the offset of its linear segments, and the
offset of the endpoints of circular arcs:
the interior of the indented circular arcs can be ignored.

\begin{corollary}\label{cor:algorithm_correct}
Algorithm~\ref{alg:decision_general} (\algdecide)~returns \YES\ if and only if there exists a
polygonal region $P$ such that $\offset(P,r)$ is
$\eps$-close to $Q$.
\end{corollary}

\begin{proof}
$Q_\eps$ is a bulged region with radius $\eps$. Therefore,
$Q_\eps^C$ is an indented region with the same radius. Since
$r>\eps$, Theorem~\ref{thm:concave-offset} implies that
$\offset(Q_\eps^C,r)$ is a bulged region with radius~$r$,
and so, $\offset(Q_\eps^C,r)^C=\inset(Q_\eps,r)=\TP$ is an
indented region with the same radius. Using $r+\eps>r$ and
applying Theorem~\ref{thm:concave-offset} once more, there
exists a polygonal region $P\subseteq\TP$ such that
$\offset(\TP,r+\eps)=\offset(P,r+\eps)$. It follows that,
if the algorithm returns \YES, there is indeed a polygonal
region~$P$ whose $r$-offset is $\eps$-close to~$Q$. If the
algorithm returns \NO, it is clear that no such polygonal
region can exist.
\end{proof}

\begin{theorem}\label{thm:complexity-statement}
Let $P$ be an indented region with radius~$r_1$ having $n$
vertices, and assume $r_2 > r_1$. Then, $\offset(P,r_2)$
has $O(n)$ vertices and it can be computed in $O(\runtime)$ time.
\end{theorem}
\begin{proof}
By Theorem~\ref{thm:concave-offset}, it suffices to consider a
polygonally bounded $P_L$ instead of $P$. We use
trapezoidal decomposition of $P$
to construct such a $P_L$ with only $O(n)$ vertices. The
Voronoi diagram of $P_L$'s vertices and (open) edges can be
computed in $O(n\log n)$ time and has size
$O(n)$~\cite{yap87}. From it, the offset polygon with the
same asymptotic complexity can be obtained in linear
time~\cite{Held91}.
\end{proof}


\begin{corollary}
Algorithm~\ref{alg:decision_general} (\algdecide)\
decides $\eps$-closeness with $O(\runtime)$ operations.
\end{corollary}
\begin{proof}
Apply Theorem~\ref{thm:complexity-statement} in each step
of Algorithm~\ref{alg:decision_general}. The fourth step 
runs in~$O(n\log n)$ time as well using a simple sweep-line algorithm.
\end{proof}

Note that $\Pi_L$, if constructed
for $\Pi$ as in the proof of Theorem~\ref{thm:concave-offset} during
step~(3) of Algorithm~\ref{alg:decision_general}, is a solution
to the deconstruction problem
if \algdecide\ returns \YES.

\section{Rational Approximation}
\label{sec:rational_approx}

A direct realization of Algorithm~\ref{alg:decision_general}
runs into difficulties
since vertices of the resulting offsets are algebraic
numbers whose degrees become high in cascaded offset
computations. We next describe two approximation variants
of Algorithm~\ref{alg:decision_general}, each producing a
certified one-sided decision by approximating all disks in
the algorithm with $k$-gons. In order to make guaranteed
statements about the exact $\eps$-approximability by
$r$-offsets, we have to approximate the disks by a
``working precision'' $\delta$ which is even smaller than
$\eps$. Recall that $D_r$ is the disk of radius $r$
centered at the origin. For $a,b\in\R$, $a<b$ define
$\annpol{a}{b}$ to be a polygon with rational vertices
whose boundary lies in the annulus $D_b\setminus D_a$.
In the approximation algorithms, every disk is replaced
with such a polygon lying inside a $\delta$-width annulus.

\ignore{
\begin{algorithm*}[ht]
\caption{One-sided approximate decision algorithms}
\label{alg:decision_app}
\captionsetup[subfloat]{textfont=normalsize, font={rm,md,sc}, labelformat=simple, labelsep=period}
\subalg[\textsc{\algdecideappinteriorfull}]{ \label{alg:decision_app_interior} \begin{minipage}{0.5\linewidth}%
\begin{compactenum}[(1)]
\item $\widehat{Q_\eps}\gets Q\oplus \widehat{D_\eps}$ with $\widehat{D_\eps}\gets \annpol{\eps-\delta}{\eps}$
\item $\widehat{\TP}\gets \left(\widehat{Q_\eps}^C\oplus\widehat{D_r}\right)^C$ with $\widehat{D_r}\gets\annpol{r}{r+\delta}$
\item $\widehat{Q'}\gets \widehat{\TP}\oplus\widehat{D_{r+\eps}}$ with $\widehat{D_{r+\eps}}\gets\annpol{r+\eps-\delta}{r+\eps}$
\item if $Q\subseteq \widehat{Q'}$, return \YES, \\
otherwise, return \UNDECIDED
\end{compactenum}
\end{minipage} }
\subalg[\textsc{\algdecideappexteriorfull}]{ \label{alg:decision_app_exterior} \begin{minipage}{0.5\linewidth}%
\begin{compactenum}[(1)]
\item $\widehat{Q_\eps}\gets Q\oplus \widehat{D_\eps}$ with $\widehat{D_\eps}\gets \annpol{\eps}{\eps+\delta}$
\item $\widehat{\TP}\gets \left(\widehat{Q_\eps}^C\oplus\widehat{D_r}\right)^C$ with $\widehat{D_r}\gets\annpol{r-\delta}{r}$
\item $\widehat{Q'}\gets \widehat{\TP}\oplus\widehat{D_{r+\eps}}$ with $\widehat{D_{r+\eps}}\gets\annpol{r+\eps}{r+\eps+\delta}$
\item if $Q\subseteq \widehat{Q'}$, return \UNDECIDED,\\
      otherwise, return \NO
\end{compactenum}
\end{minipage} }
\end{algorithm*}
}
\myparagraph{Interior approximation}
In the first part of our algorithm, we ensure that the
final approximation of $Q'$ (see line~(3) of
Algorithm~\ref{alg:decision_general}), called
$\widehat{Q'}$ , will be a subset of the exact $Q'$. We
achieve this by approximating $D_s$ by
$\annpol{s-\delta}{s}$ when an offset is computed; and by
approximating $D_s$ by $\annpol{s}{s+\delta}$ when an inset
is computed; see 
Algorithm~\ref{alg:decision_app_interior}.

\begin{algorithm}[h]
\caption{\textsc{\algdecideappinteriorfull}}
\begin{compactenum}[(1)]
\item $\widehat{Q_\eps}\gets Q\oplus \widehat{D_\eps}$ with $\widehat{D_\eps}\gets \annpol{\eps-\delta}{\eps}$
\item $\widehat{\TP}\gets \left(\widehat{Q_\eps}^C\oplus\widehat{D_r}\right)^C$ with $\widehat{D_r}\gets\annpol{r}{r+\delta}$
\item $\widehat{Q'}\gets \widehat{\TP}\oplus\widehat{D_{r+\eps}}$ with $\widehat{D_{r+\eps}}\gets\annpol{r+\eps-\delta}{r+\eps}$
\item if $Q\subseteq \widehat{Q'}$, return \YES, \\
	otherwise, return \UNDECIDED
\end{compactenum}
\label{alg:decision_app_interior}
\end{algorithm}
\begin{lemma}\label{lem:one_side_correctness}
If \algdecideappinteriorfull returns \YES, then
\algdecidefull\ returns \YES\ as well, which means that
there exists a polygonal region $P$ such that
$\offset(P,r)$ is $\eps$-close to $Q$. In particular,
$P:=\widehat{\TP}$ is a solution to the deconstruction problem.
\end{lemma}
\begin{proof}
Compare the execution of
Algorithm~\ref{alg:decision_app_interior} with the 
corresponding call of its exact version,
Algorithm~\ref{alg:decision_general}. It is
straight-forward to check that for any $\delta$,
$\widehat{Q_\eps}\subset Q_\eps$,
$\widehat{\TP}\subset\TP$, and $\widehat{Q'}\subset Q'$.
The last inclusion shows that if $Q\subseteq\widehat{Q'}$,
also $Q\subseteq Q'$.
\end{proof}

\begin{definition}\label{def:criteps}
For fixed $Q$ and $r$, define
$\criteps:=\inf\{\eps\mid \algdecidefull \text{ returns \YES}\}.$
\end{definition}

Note that $\criteps\in[0,r]$, and that \algdecidefull\ returns
\YES\ for every $\eps\geq\criteps$ and returns \NO\
for every $\eps<\criteps$. We do not have a way to compute 
$\criteps$ exactly. However, we show next that
\algdecideappinteriorfull returns \YES\ for every
$\eps>\criteps$ for $\delta$ small enough, and that the
required precision $\delta$ is proportional to the distance
of $\eps$ to $\criteps$.

\begin{theorem}\label{thm:precision_quality_interior}
Let $\eps>\criteps$, and $\delta<\frac{\eps-\criteps}{2}$.
Then, \algdecideappinteriorfull returns \YES.
\end{theorem}

\textsc{Proof.}
Let $\eps_0$ be such that
$\criteps<\eps_0<\eps_0+2\delta\leq\eps$. Let $Q_{\eps_0}$,
$\TP$ and $Q'$ denote the intermediate results of
\algdecide$(Q,r, \eps_0)$ and let $\widehat{Q_\eps}$, $\widehat{\TP}$,
$\widehat{Q'}$ denote the intermediate results of \algdecideappinteriorfull. 
By the choice of $\eps_0$, \YES\
is returned, and thus $Q\subseteq Q'$. The theorem follows from
$Q'\subseteq \widehat{Q'}$, which we prove in three substeps:
\begin{itemize}
\item [(1) $\offset(Q_{\eps_0},\delta)\subseteq \widehat{Q_\eps}$:]
Indeed,\\
$$\offset(Q_{\eps_0},\delta) =  \offset(Q,\eps_0+\delta)
                           \subseteq  \offset(Q,\eps-\delta)
                           \subset  Q\oplus\widehat{D_\eps}
                           =  \widehat{Q_\eps}.$$ƒ
\item [(2) $\TP\subseteq\widehat{\TP}$:] Starting with (1), we obtain
\begin{eqnarray*}
&&\offset(Q_{\eps_0},\delta)\subseteq \widehat{Q_\eps}\\
&\Rightarrow& \offset(Q_{\eps_0},\delta)^C\oplus
D_{r+\delta}\supseteq \widehat{Q_\eps}^C\oplus\widehat{D_r}\\
&\Rightarrow&\inset(\offset(Q_{\eps_0},\delta),
r+\delta)\subseteq \widehat{\TP}\;.
\end{eqnarray*}
We use the general fact
$\inset(\offset(A,r),r)\supseteq A$ to obtain:
$$\inset(\offset(Q_{\eps_0},\delta), r+\delta)
=\inset(\inset(\offset(Q_{\eps_0},\delta),\delta),r)
\supseteq\inset(Q_{\eps_0},r)=\TP.$$
\item [(3) $Q'\subseteq\widehat{Q'}$:] Using (2), we have that
$$\offset(\TP,r+\eps-\delta)=\TP\oplus D_{r+\eps-\delta}\subseteq \widehat{\TP}\oplus\widehat{D_{r+\eps}}=\widehat{Q'}.$$
Note that $r+\eps-\delta>r+\eps_0$, and therefore,
$\offset(\TP,r+\eps-\delta)\supset\offset(\TP,r+\eps_0)=Q'$.\qed
\end{itemize}

\myparagraph{Exterior approximation}
In 
Algorithm~\ref{alg:decision_app_exterior}, we ensure that
$\widehat{Q'}$ becomes a superset of the exact $Q'$ by
appropriately choosing approximate disks. Specifically, we
approximate $D_s$ by $\annpol{s}{s+\delta}$ when an offset
is computed, and $D_s$ by $\annpol{s-\delta}{s}$ when an
inset is computed.
Not surprisingly, we
get a certified answer in the other direction, and a certified answer
is guaranteed when $\delta$ is sufficiently small. The proofs of the 
following two statements are similar to Lemma~\ref{lem:one_side_correctness} 
and Theorem~\ref{thm:precision_quality_interior}
and thus omitted.
\begin{algorithm}[h]
\caption{\textsc{\algdecideappexteriorfull}}
\begin{compactenum}[(1)]
\item $\widehat{Q_\eps}\gets Q\oplus \widehat{D_\eps}$ with $\widehat{D_\eps}\gets \annpol{\eps}{\eps+\delta}$
\item $\widehat{\TP}\gets \left(\widehat{Q_\eps}^C\oplus\widehat{D_r}\right)^C$ with $\widehat{D_r}\gets\annpol{r-\delta}{r}$
\item $\widehat{Q'}\gets \widehat{\TP}\oplus\widehat{D_{r+\eps}}$ with $\widehat{D_{r+\eps}}\gets\annpol{r+\eps}{r+\eps+\delta}$
\item if $Q\subseteq \widehat{Q'}$, return \UNDECIDED,\\
      otherwise, return \NO
\end{compactenum}
\label{alg:decision_app_exterior}
\end{algorithm}
\begin{lemma}\label{lem:one_side_guarantee_exterior}
If \algdecideappexteriorfull returns~\NO, then
\algdecidefull\ returns~\NO\ as well, which means that
there exists no polygonal region $P$ such that
$\offset(P,r)$ is $\eps$-close to $Q$.
\end{lemma}

\begin{theorem}\label{thm:precision_quality_exterior}
Let $\eps<\criteps$ and $\delta<\frac{\criteps-\eps}{2}$.
Then, \algdecideappinteriorfull returns~\NO.
\end{theorem}

In combination with Theorem~\ref{thm:precision_quality_interior}, it follows
that the exact answer can always be found for 
$\delta<\Delta:=\frac{|\eps-\criteps|}{2}$ by combining \algdecideappexterior\
and \algdecideappinterior.
We display the complete rational approximation algorithm\footnote{As in the \algdecide case 
for $\eps = 0$ the return value is $\NO$, and for $\eps = r$ it is $\YES$ with $P = Q$.}
for later reference:

\begin{algorithm}[h]
\caption{\textsc{\algdecideappfull}}
\begin{compactenum}[(1)]
\item if \algdecideappinteriorfull\ $=$ \YES, return \YES
\item if \algdecideappexteriorfull\ $=$ \NO, return \NO
\item Otherwise, return \UNDECIDED
\end{compactenum}
\label{alg:decision_app_full}
\end{algorithm}

\begin{wrapfigure}[9]{r}{4.0cm}
\resizebox{4.0cm}{!}{\includegraphics{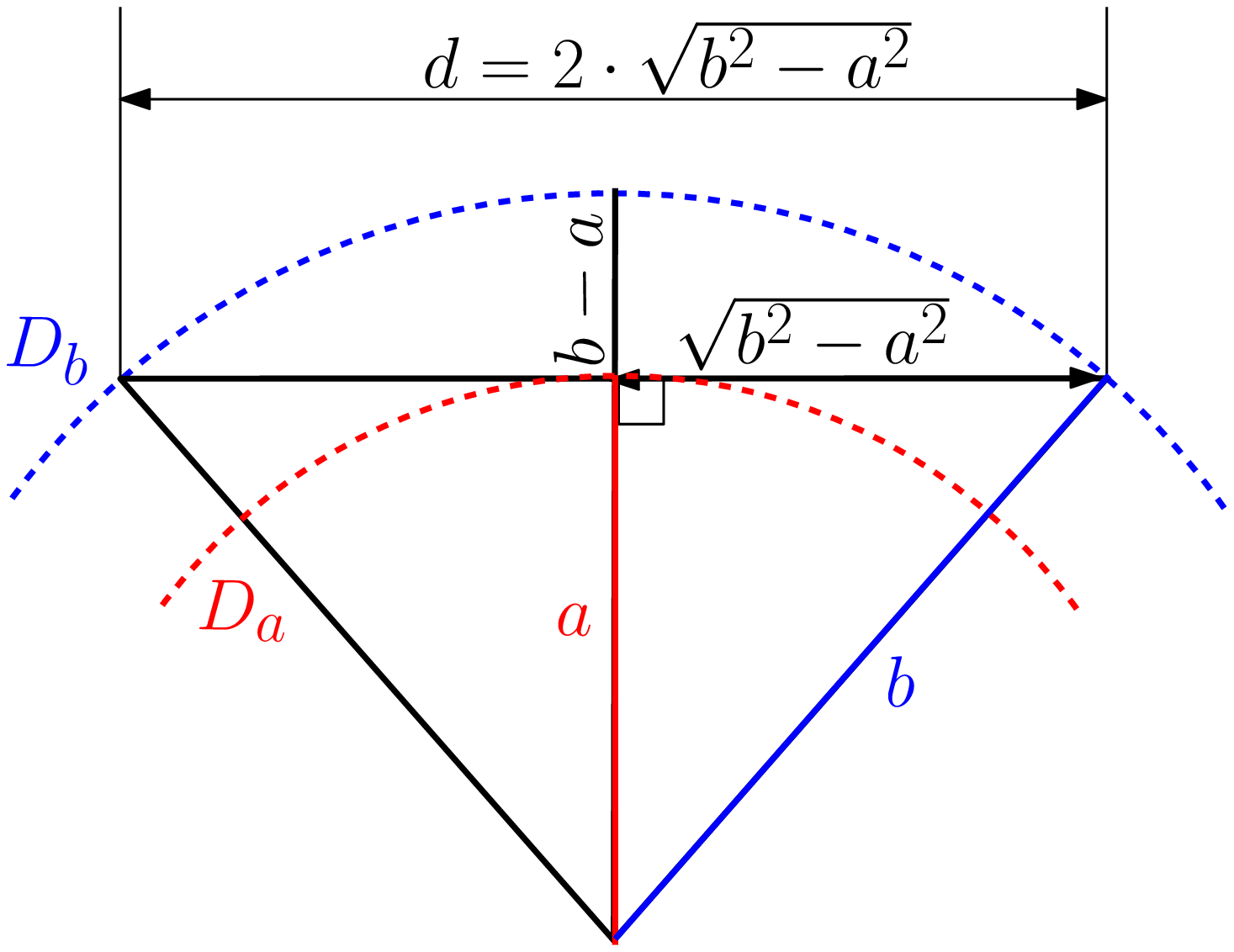}}
\end{wrapfigure}
\myparagraph{Complexity analysis}
The main task is to bound the number of vertices of $\annpol{a}{b}$.
We will create a $\annpol{a}{b}$ with the additional property 
that all vertices lie on $\partial D_b$.
As depicted on the right, two such points on $D_b$ are connected by a chord
of the boundary circle that does not intersect $D_a$ if and only if
the angle induced by the two points is at most $\psi:=2\arccos\frac{a}{b}$,
or equivalently, the length of the chord is less than $2\sqrt{b^2-a^2}$.
Note that we need at least $\frac{2\pi}{\psi}$ points on $\partial D_b$ for a
valid $\annpol{a}{b}$, and $\frac{2\pi}{\psi}\in\Theta(\sqrt{\frac{b}{b-a}})$
as easily shown by L'Hopital's rule.

Rational points on $\partial D_b$ can be constructed 
for an arbitrary $t\in\Q$ as
$Q_t:=(b\frac{1-t^2}{1+t^2},b\frac{2t}{1+t^2})$~\cite{142726}. 
For some positive
$z\in\Z$, we define $P_i:=Q_{i/z}$ for $i=0,\ldots,z$. 
\ignore{
The next lemma follows from elementary geometry, in particular
the \emph{Angle Bisector
Theorem}.\footnote{\url{http://en.wikipedia.org/wiki/Angle_bisector_theorem}}
}
\begin{lemma}\label{lem:chords}
For every $i=1,\ldots,z-1$, the chord $P_{i-1}P_{i}$ is
longer than the chord $P_{i}P_{i+1}$. In particular, the
length of each chord is bounded by the length of $P_0P_1$
which is shorter than $\frac{2b}{z}$.
\end{lemma}

\begin{wrapfigure}[8]{r}{5cm}
\resizebox{5.0cm}{!}{\label{fig:chords}\includegraphics{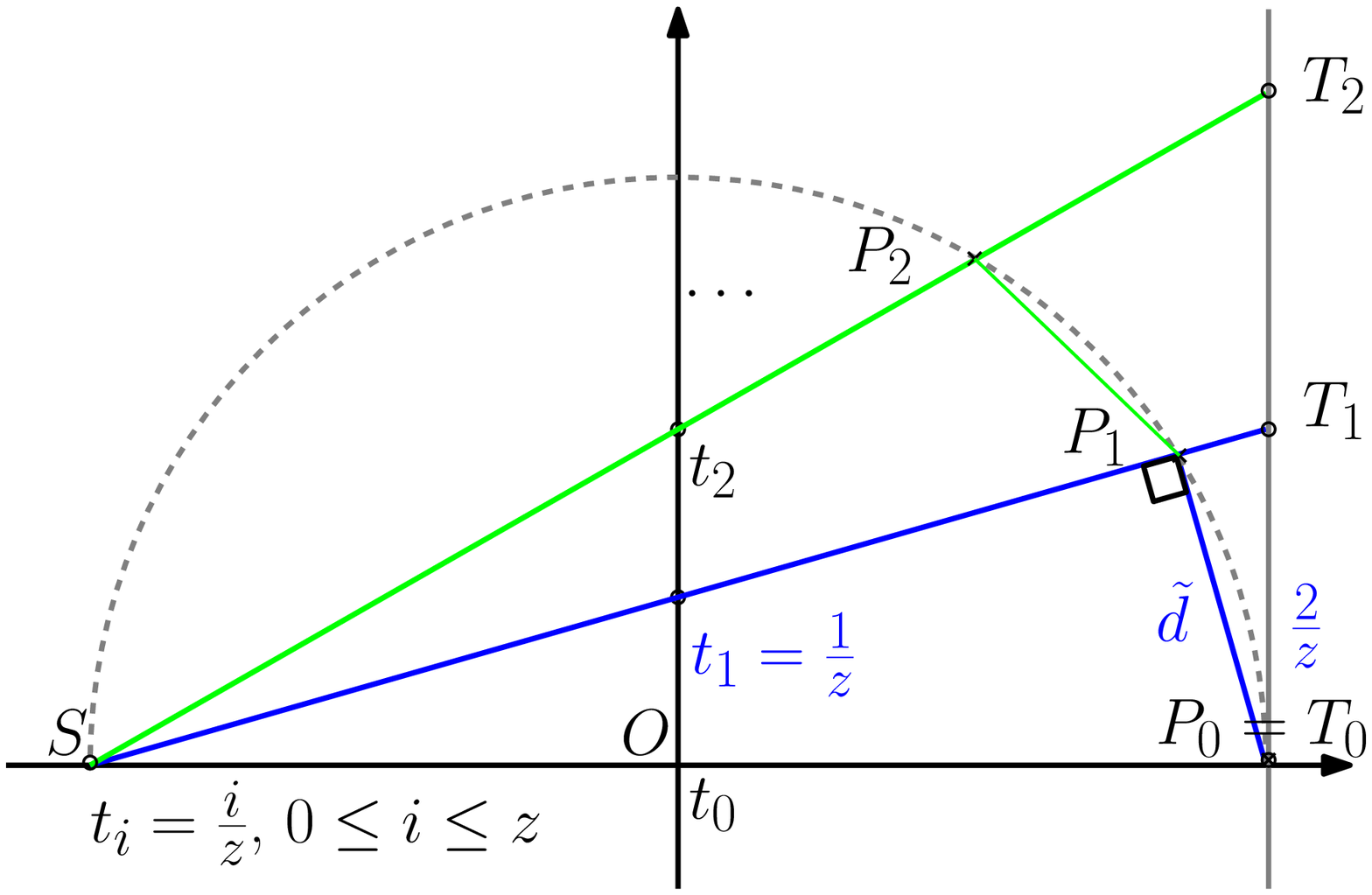}}
\end{wrapfigure}
\textsc{Proof.}
W.l.o.g., we assume $b=1$ for the proof, since the chord
length scales proportionally when scaling the circle by a
factor of~$b$.
The point $Q_t=(\frac{1-t^2}{1+t^2},\frac{2t}{1+t^2})$ can
be constructed geometrically as the intersection point of
$\partial D_b$ with the line $\ell_t$ through $S=(-1,0)$
and slope $t$ (see the figure on the right). 
In particular,
the line $SP_i$ has slope $\frac{i}{z}$; we let $T_i$
denote the intersection point of that line with the line
$x=1$. We observe that the segment $T_iT_{i+1}$ has length
$\frac{2}{z}$, and that $ST_i<ST_{i+1}$ for
$i=0,\ldots,z-1$.

We are showing next that the chord $P_{i-1}P_i$ is longer
than $P_iP_{i+1}$. For that, we consider the triangle
$SD_{i-1}D_{i+1}$, and its bisector at $S$. This bisector
intersects the line $x=1$ at some point $B$. By the Angle
Bisector theorem,
$B$ divides the segment $T_{i-1}T_{i+1}$ proportionally to
the corresponding triangle sides, that is,
$\ddfrac{ST_{i-1}}{ST_{i+1}} =
\ddfrac{BT_{i-1}}{BT_{i+1}}$. Because the left-hand side is
smaller than $1$, it follows that $BT_{i-1}$ is shorter
than $BT_{i+1}$. Therefore, $B$ lies below $T_i$, and
therefore, the angle $\alpha_{i-1}=\angle
T_{i-1}ST_i=\angle P_{i-1}SP_i$ is larger than
$\alpha_{i}=\angle T_{i}ST_{i+1}=\angle P_{i}SP_{i+1}$. But
the chord lengths $P_{i-1}P_i$ and $P_iP_{i+1}$ are defined
by $2\sin(\alpha_{i-1})$ and $2\sin(\alpha_{i})$,
respectively, which proves that the chord lengths are
indeed decreasing.

Finally, by Thales' theorem, the triangle $SP_0P_1$ has a right angle at $P_1$.
Therefore, the longest chord $P_0P_1$ is shorter than the segment $T_0T_1$, which has
length $\frac{2}{z}$.\qed

Note that all $P_i$'s lie in the first quadrant of the
plane and that $P_0:=(b,0)$ and $P_z:=(0,b)$. Therefore, we
can subdivide the other three quarters of the circle
symmetrically such that the length of each chord is bounded
by $\frac{2b}{z}$, using $4z$ vertices altogether. To
compute a valid $\annpol{a}{b}$, it suffices to choose
$z$ such that $\frac{2b}{z}\leq 2\sqrt{b^2-a^2}$, that is
$z \geq \sqrt{\frac{b^2}{b^2-a^2}}$. We choose
$z_0:=\left\lceil \sqrt{\frac{b}{b-a}}\right\rceil$,
indeed, since $0<a<b$, we have that
$z_0\geq\sqrt{\frac{b}{b-a}}>\sqrt{\frac{b}{b-a}\cdot\frac{b}{b+a}}=\sqrt{\frac{b^2}{b^2-a^2}}$.
As stated above, we need at least
$\Omega(\sqrt{\frac{b}{b-a}})$ points, so $z_0$ is an
asymptotically optimal choice. We summarize the result
\begin{lemma}\label{lem:annulus_bound}
For $a<b$, a polygonal region $\annpol{a}{b}$ as above with $O(\sqrt{\frac{b}{b-a}})$ (rational) points
can be computed using $O(\sqrt{\frac{b}{b-a}})$ arithmetic operations.
\end{lemma}
The Minkowski sum of an arbitrary polygonal region with $n$
vertices and a convex polygonal region with $k$ vertices
has complexity $O(kn)$ and it can be computed in
$O(nk\log^2(nk))$ operations  by a simple
divide-and-conquer approach, using a sweep line algorithm in
the conquer step~\cite{klps-ujrcf-86}. Using generalized
Voronoi diagrams where the distance is based on the convex
summand of the Minkowski sum operation~\cite{ls-pptmc-87},
we obtain an improved algorithm, which requires only
$O(kn\log (kn))$ operations. In combination with
Lemma~\ref{lem:annulus_bound}, this leads to the following
complexity bound for the two approximation algorithms.
\begin{theorem}\label{thm:rational_complexity}
Algorithm \algdecideapp requires
$$O(n\frac{r}{\delta}\sqrt{\frac{\eps}{\delta}}\cdot \log(n\frac{r}{\delta}\sqrt{\frac{\eps}{\delta}}))$$
arithmetic operations with rational numbers.
\end{theorem}
We remark that the $O(n\log n)$ bound for \algdecide\ refers to operations with real numbers instead.

We have implemented the algorithms \algdecideappinterior and \algdecideappexterior
using exact rational arithmetic using the \cgal%
\footnote{The Computational Geometry Algorithms, \url{www.cgal.org}}
packages for polygons~\cite{cgal:gw-p2-10b}, 
Minkowski sums%
\ignore{
\footnote{Notice that we use a
convolution-cycles based implementation of Minkowski sums,
which is known to perform very well in practice
and in particular it was experimentally shown to perform
better than divide-and-conquer based implementations described
in Section~\ref{sec:rational_approx} on
many inputs~\cite{w-eacop-07}.}
}
~\cite{cgal:w-rms2-10b} and Boolean
set operations~\cite{cgal:fwzh-rbso2-10b}.
We demonstrate the execution of our software on two examples
in Figures~\ref{fig:approximability_parameters}
and~\ref{fig:real_yes}.

\def\YesWheel{{%
\includegraphics[width=0.23\textwidth]{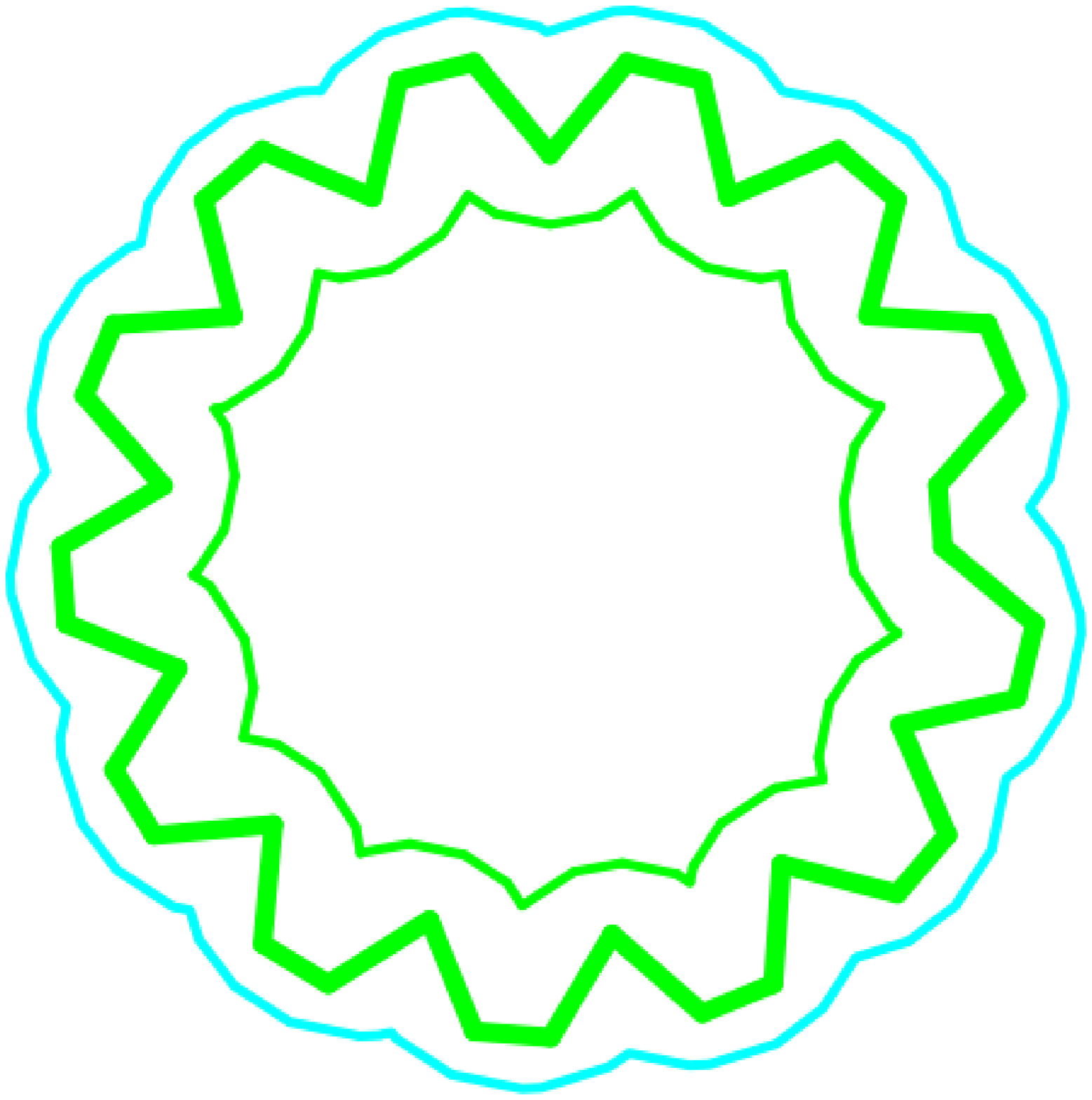}
}}
\def\NoWheel{{%
\includegraphics[width=0.23\textwidth]{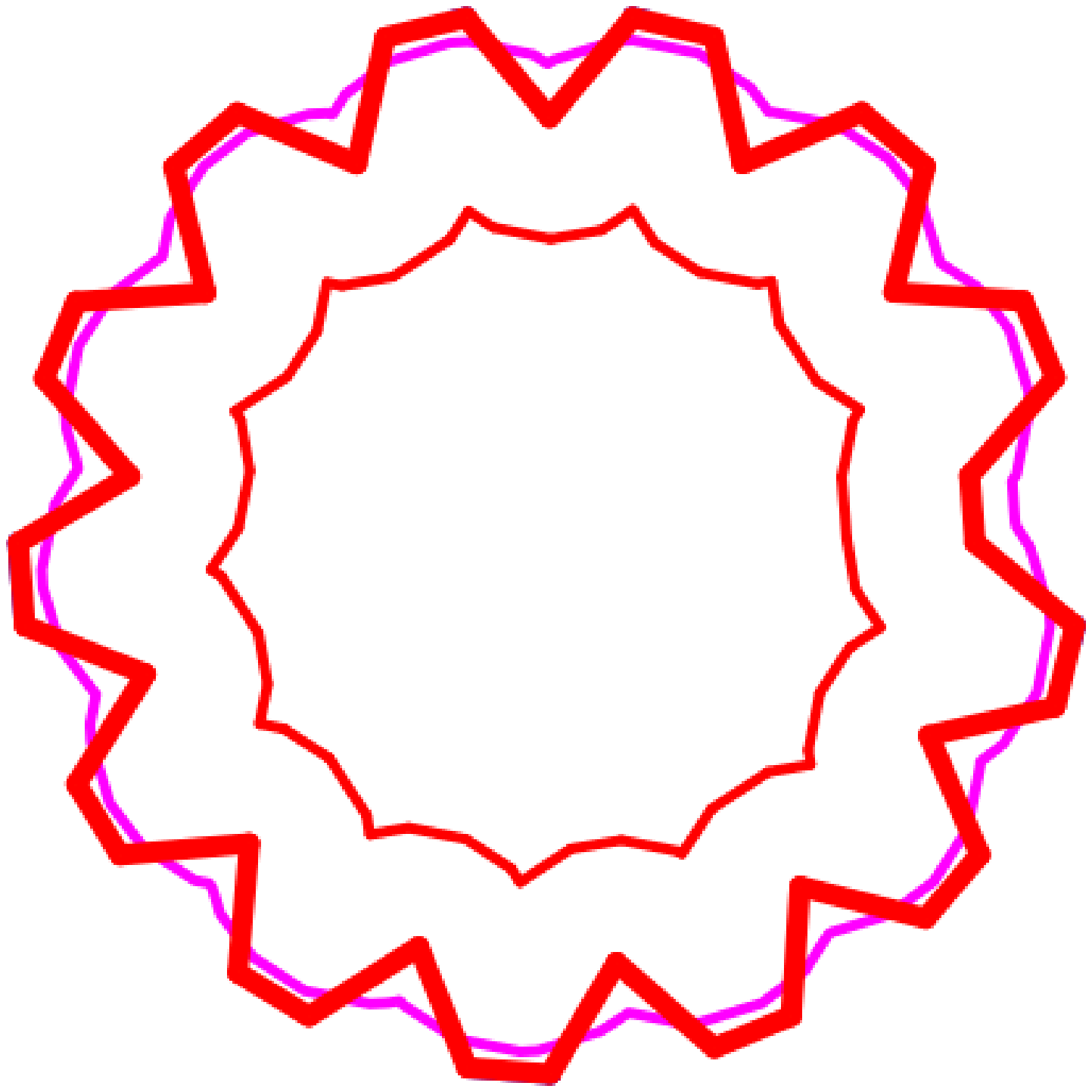}
}}
\def\UndWheel{{%
\includegraphics[width=0.23\textwidth]{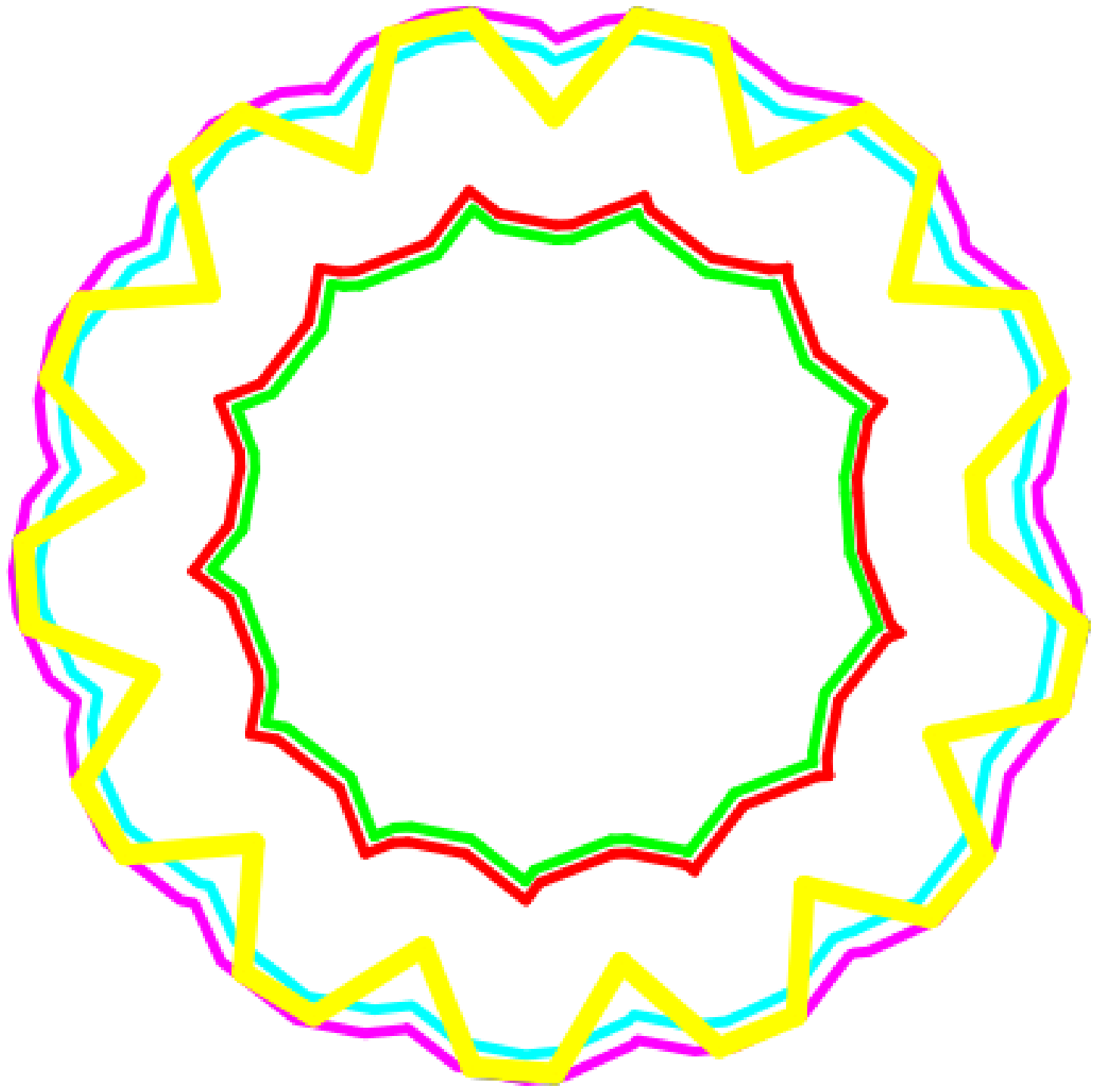}
}}
\def\UndNoWheel{{%
\includegraphics[width=0.23\textwidth]{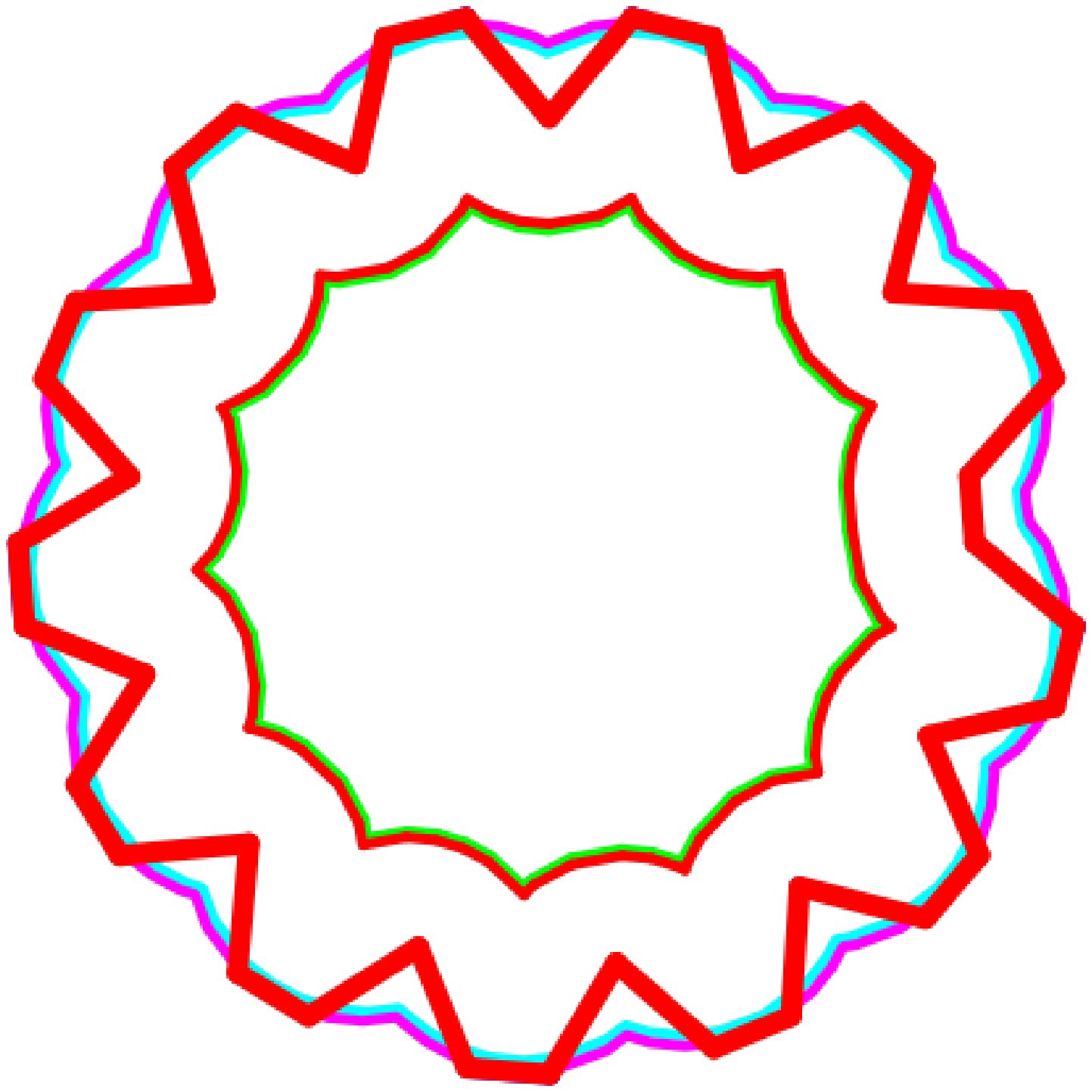}
}}
\def\YesWheelM{{%
\includegraphics[width=0.25\textwidth]{wheel_y_1_3_c_abs.eps}
}}
\def\NoWheelM{{%
\includegraphics[width=0.25\textwidth]{wheel_n_1_9_c.eps}
}}
\def\UndWheelM{{%
\includegraphics[width=0.25\textwidth]{wheel_u_1_6_1_4_c.eps}
}}
\def\UndNoWheelM{{%
\includegraphics[width=0.25\textwidth]{wheel_n_1_6_1_10_c.eps}
}}
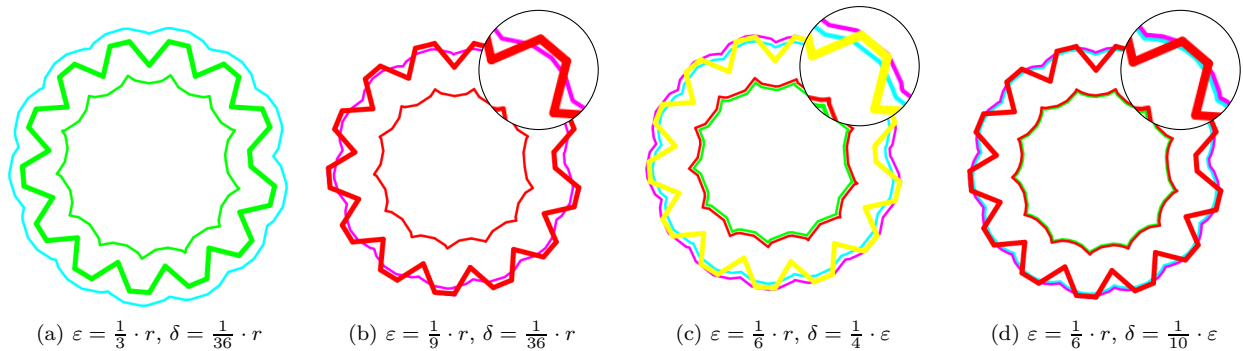
\begin{figure*}[]
\subfloat[$\eps = \frac{1}{3} \cdot r$, $\delta = \frac{1}{36} \cdot r$]{
\label{fig:yes_eps}
\YesWheel
}
\subfloat[$\eps = \frac{1}{9} \cdot r$, $\delta = \frac{1}{36} \cdot r$]{
\label{fig:no_eps}
\NoWheel
\begin{pspicture}(1.9, -2.0)(2,2)
\newpsstyle{NoShadow}{ shadow=false }
\PstLens[LensHandle=false, LensSize=0.8, LensMagnification=1.6, LensStyleGlass=NoShadow](1.0,1.2){\NoWheelM}
\end{pspicture}
}
\subfloat[$\eps = \frac{1}{6} \cdot r$, $\delta = \frac{1}{4} \cdot \eps$]{
\label{fig:und_delta}
\UndWheel
\begin{pspicture}(1.9, -2.0)(2,2)
\newpsstyle{NoShadow}{ shadow=false }
\PstLens[LensHandle=false, LensSize=0.8, LensMagnification=1.6, LensStyleGlass=NoShadow](1.0,1.25){\UndWheelM}
\end{pspicture}
}
\subfloat[$\eps = \frac{1}{6} \cdot r$, $\delta = \frac{1}{10} \cdot \eps$]{
\label{fig:no_delta}
\UndNoWheel
\begin{pspicture}(1.9, -2.0)(2,2)
\newpsstyle{NoShadow}{ shadow=false }
\PstLens[LensHandle=false, LensSize=0.8, LensMagnification=1.6, LensStyleGlass=NoShadow](1.0,1.2){\UndNoWheelM}
\end{pspicture}
}
{
   \caption{Dependency of the algorithm outcome on $\eps$ and $\delta$:
  The input polygon (wheel) appears in bold
  line. It is colored according to its approximability with
  the given parameters: green for \YES, red for \NO\ and
  yellow for \UNDECIDED. The inset polygon $\widehat{TP}$
  and its approximate $(r+\eps)$-offset $\widehat{Q'}$ 
  are drawn in green and cyan respectively. Their
  outer-approximation counterparts are drawn in red and magenta.
  Figures~\protect\subref{fig:yes_eps} and~\protect\subref{fig:no_eps}
  demonstrate how  when $\eps$ is tightened from $\frac{1}{3} \cdot
  r$ to $\frac{1}{9} \cdot r$, with the same $r$ and
  $\delta$, the decision result changes from \YES\ to \NO. The
  green polygon inside the input polygon in~\protect\subref{fig:yes_eps}
  is a possible $r$-offset solution. The magnification in~\protect\subref{fig:no_eps}
  highlights the area of the input polygon that does not fit inside the
  outer $\delta$-approximation (in magenta) of maximal possible $(r+\eps)$-offset.
  Figures~\protect\subref{fig:und_delta} and~\protect\subref{fig:no_delta} show how
  when $\delta$ is decreased from $\frac{1}{4} \cdot \eps$ to
  $\frac{1}{10} \cdot \eps$, for the same $r$ and $\eps$,
  the decision result changes from \UNDECIDED\ to \NO, namely in the
  latter case the algorithm is able to produce a
  certified negative answer.} \label{fig:approximability_parameters}
}
\end{figure*}

\def\Kazahstan{{%
\includegraphics[width=12.0cm]{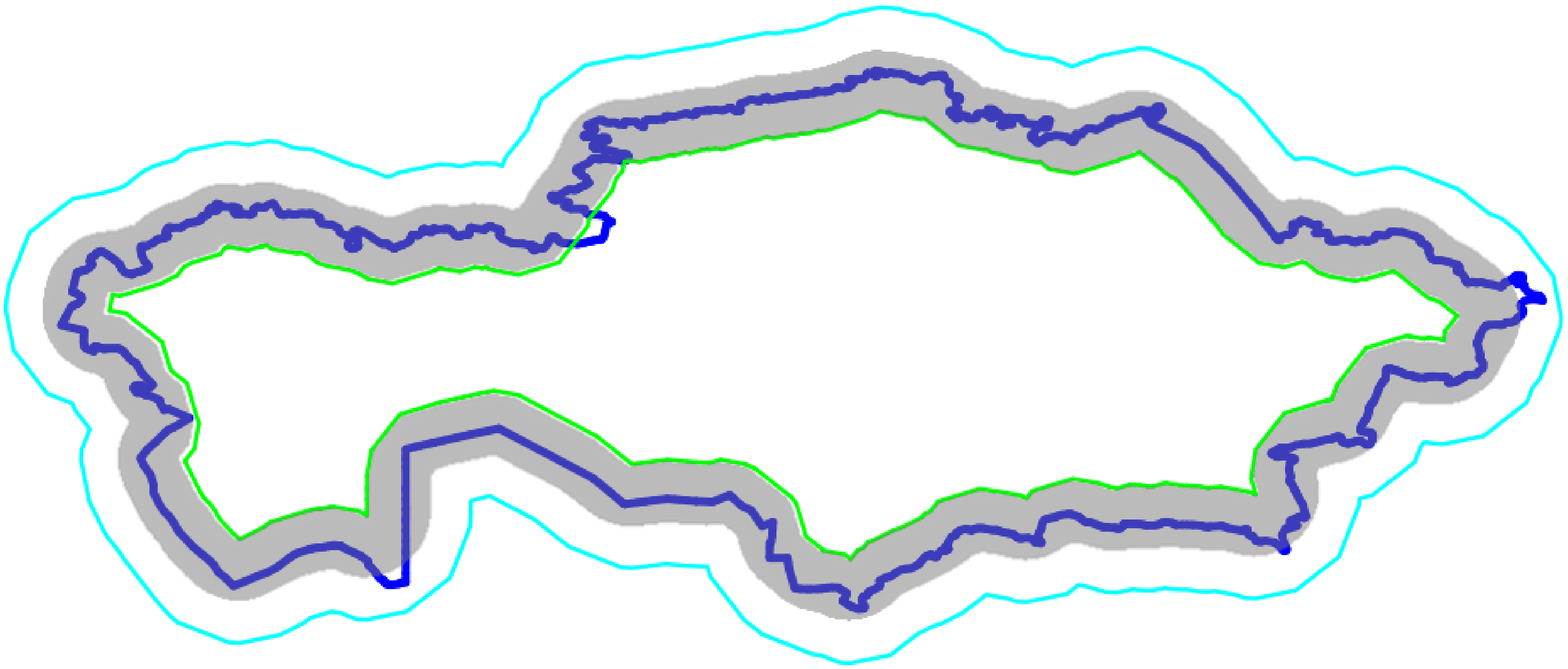}
}}
\def\KazahstanWithLens{{%
\includegraphics[width=10.0cm]{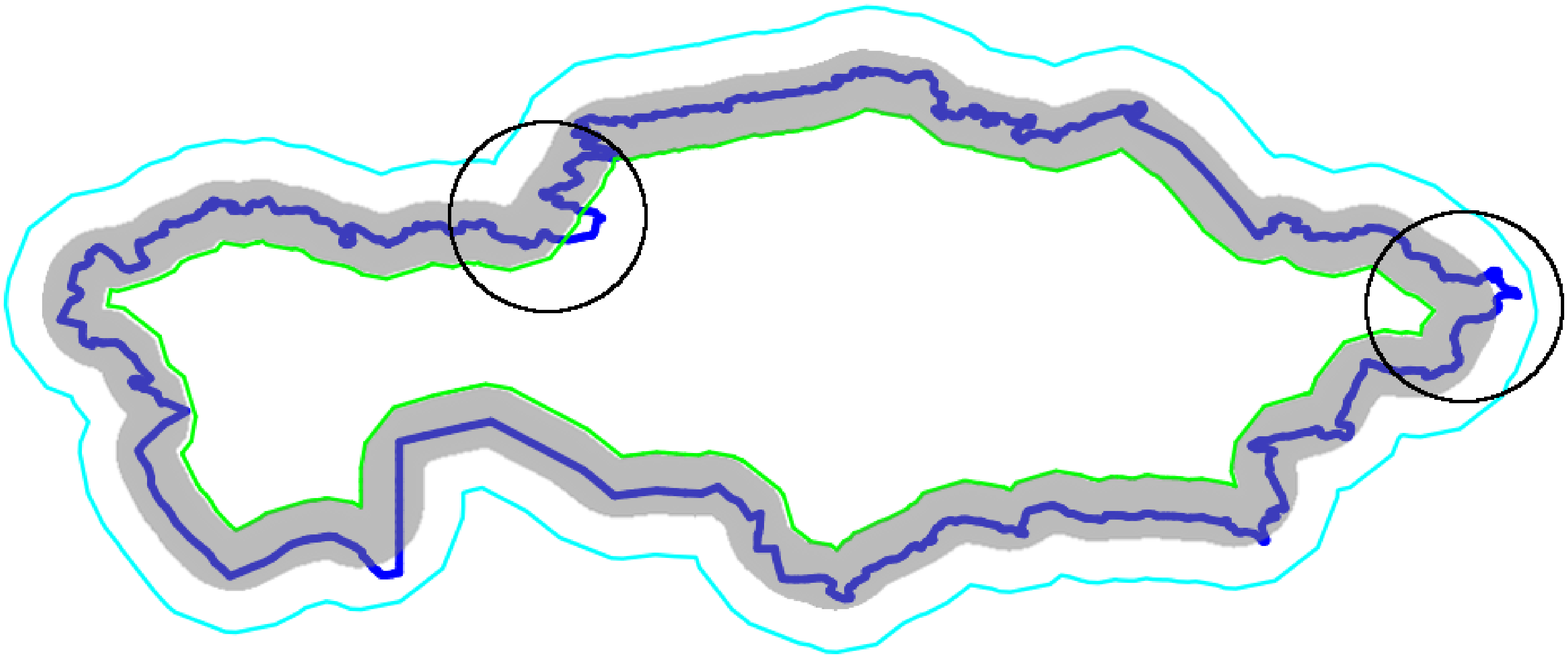} 
}}
\begin{figure*}[]
  \begin{pspicture}(-3.0,-1.8) 
  \psset{unit=1cm}
  \PstLens[LensHandle=false, LensSize=1.2, LensMagnification=1.5](-1.7,0.75){\Kazahstan} 
  \end{pspicture}
  \KazahstanWithLens
  \begin{pspicture}(3.5,-2.5)(3.5,1) 
  \psset{unit=1cm}
  \PstLens[LensHandle=false, LensSize=1.2, LensMagnification=1.5](5.0,0.15){\Kazahstan} 
  \end{pspicture}
  \caption{A map of Kazakhstan, represented as a polygon
  $Q$ (in bold blue) with 1881 vertices, is approximable
  for $\eps = \frac{1}{2} \cdot r$ and $\delta =
  \frac{1}{8} \cdot \eps$. A solution polygon $P$ (in
  green) has 335 vertices. Offset($P$,$r$) (shown as
  lightly-shaded gray $r$-strip around $P$) is inside the $\eps$-offset
  of the input $Q$ by construction. The $\delta$-approximation of the 
  $\eps$-offset of Offset($P$,$r$) (as computed in line~(3)
  of \algdecideappinteriorfull) is drawn in cyan and has 261
  vertices. Since the cyan polygon contains $Q$, the
  Offset($P$,$r$) and $Q$ have Hausdorff distance of at most
  $\eps$, that is, $Q$ is approximable and $P$ is a solution. 
  Approximability computation took 3.868 seconds in
  this case on a 3GHz Intel Dual Core processor. 
  The magnification on the left
  highlights some cavities in the input polygon that have
  no effect on the Hausdorff distance within this tolerance $\eps$. 
  The magnification on the right demonstrates a sharp end 
  that would prevent $Q$'s approximability with a tighter $\eps$.}
  \label{fig:real_yes}
  
\end{figure*}
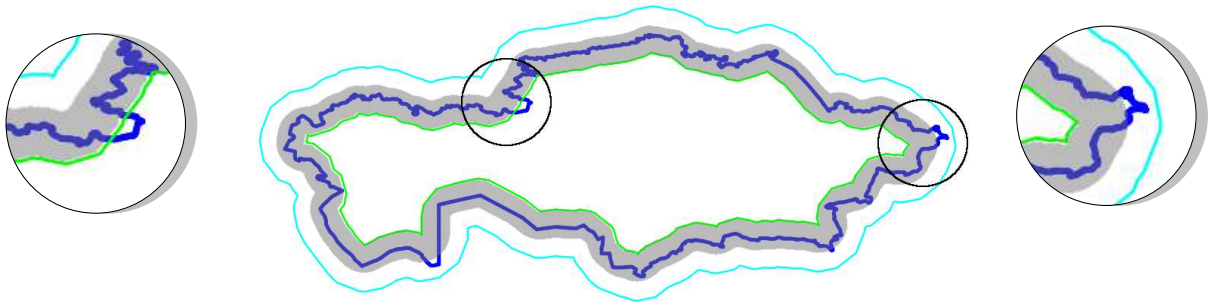

%

\section{Searching $\eps$ and $r$}
\label{sec:search}

So far, we have assumed that both $r$ and $\eps$ are given as input parameters,
and we posed the question of deconstructing a polygon with respect to these
parameters.
We now investigate three variants where $r$ and/or $\eps$ are unknown. 
Specifically, we ask, for some input polygon~$Q$:
\begin{enumerate} 
\item Given $r$, what is $\criteps$, the infimum of all $\eps$-values 
      such that the deconstruction problem has a solution (compare Definition~\ref{def:criteps})?
\item Given $\eps$, what is the set of radii for which the deconstruction problem has a solution?
\item Given neither $r$ nor $\eps$, how to choose them in a ``reasonable'' way to
      obtain a solution?
\end{enumerate}
Whereas the first two questions are formally posed, 
the third one is of a rather heuristic nature.
In all three cases, we also ask for computing some polygonal shape $P$
that approximates the solution of
the deconstruction problem for the given set of parameters. 

We discuss the posed questions in the remainder of this section. 
Our main tool will be the decision algorithm for fixed $r$ and $\eps$
as described earlier.
Because we aim for a practical algorithm, 
we formulate our approach using the rational approximation algorithm
from Section~\ref{sec:rational_approx}.
We have implemented the proposed algorithms;
the example at the end of this section has been produced
with our implementation.

\myparagraph{Searching for $\criteps$}
If we use the exact decision procedure \algdecide, it is straight-forward
to approximate $\criteps$ to arbitrary precision $\Delta$ employing binary search:
Start with the interval $[0,r]$ and choose $\eps$ as the midpoint of the interval.
If \algdecidefull\ returns \YES, recurse on the left subinterval, otherwise,
on the right one.
Obviously, the interval width is halved in every step, so $O(\log(\frac{r}{\Delta}))$ 
steps are necessary. 
Let $\critepsappd$ denote the $\criteps$ approximation, s.t. $\critepsappd - \criteps \leq \Delta$.
We demonstrate next that we can achieve the same approximation and produce 
with it a solution to the deconstruction problem using the rational 
approximation version \algdecideapp. 

Let $|I|$ denote the width of $I$ henceforth on and consider the pseudocode
given in Algorithm~\ref{alg:search_criteps}. It computes an interval
$I$ of width at most $\Delta$ that contains $\criteps$. 

\begin{algorithm}[H]
\caption{\textsc{\algsearchepsappfull}}
\begin{compactenum}[(1)]
\item $I\gets [0,r]$
\item \textbf{while} $|I|> \Delta$ \textbf{do}
\item \hspace{0.3cm} $\epsno\gets \text{left endpoint of }I$, $\epsyes\gets\text{right endpoint of }I$ 
\item \hspace{0.3cm} $\epsmid \gets \frac{\epsno + \epsyes}{2}$, $\delta \gets \frac{|I|}{8}$
\item \hspace{0.3cm}$res \gets \algdecideapp(Q, r, \epsmid, \delta)$
\item \hspace{0.3cm}if $res = \YES$ then $I \gets [\epsno,\epsmid]$
\item \hspace{0.3cm}otherwise, if $res = \NO$, then $I \gets [\epsmid,\epsyes]$
\item \hspace{0.3cm}otherwise, ($res = \UNDECIDED$), then $I \gets [\epsmid-\frac{|I|}{4},\epsmid + \frac{|I|}{4}]$
\item \textbf{end while}
\item return $I$
\end{compactenum}
\label{alg:search_criteps}
\end{algorithm}

We prove the invariant that $\criteps\in I$ 
after each iteration of the while-loop,
implying correctness of the whole algorithm.
Trivially, $\criteps\in [0,r]$,
and the invariant is obviously maintained
if $\algdecideapp(Q, r, \epsmid, \delta)$ returns \YES\ or \NO. 
For the case of \UNDECIDED, recall that \algdecideapp\ is a combination of the two one-sided
approximation algorithms \algdecideappexterior\ and \algdecideappinterior, and both
returned \UNDECIDED. Theorem~\ref{thm:precision_quality_interior} and 
Theorem~\ref{thm:precision_quality_exterior} imply therefore that 
$$\frac{|I|}{8}\geq |\frac{\epsmid-\criteps}{2}|.$$
It follows that 
$\criteps \in [\epsmid-\frac{|I|}{4},\epsmid+\frac{|I|}{4}]$
which proves that the invariant is maintained also in this case. 

We next compute a solution $P$ for the deconstruction problem for 
$Q$, $r$ and~$\critepsappd$. 
Recall that if \algdecideapp\ returns
\YES, the algorithm computes a solution for the deconstruction problem as a by-product. 
Let $I\gets\algsearchepsappfull$ be the approximation interval for $\criteps$
and let $\epsyes$ denote the right endpoint of $I$, that is $\criteps \leq \epsyes$.
We call $\algdecideapp(Q, r, \epsyes, \frac{|I|}{4})$. If the result is \YES,
then $\critepsappd = \epsyes$ and the polygon computed by $\algdecideapp$ is a solution.
Otherwise let us choose $\critepsappd = \epsyes + \frac{\Delta}{2}$ and 
produce a solution by calling $\algdecideapp(Q, r, \critepsappd, \frac{\Delta}{8})$. 
Since the result of $\algdecideapp(Q, r, \epsyes, \frac{|I|}{4})$ was \UNDECIDED\ 
we conclude from Theorem~\ref{thm:precision_quality_interior} 
that $\criteps\geq\epsyes-\frac{|I|}{2}$, that is $\critepsappd = \epsyes + \frac{\Delta}{2}$ 
is indeed $\Delta$-approximation of $\criteps$. The call to 
$\algdecideapp(Q, r, \critepsappd, \frac{\Delta}{8})$ is bound to yield \YES\ 
because if it returned \UNDECIDED, we would have that
$\criteps \geq \critepsappd - 2\frac{\Delta}{8} = \epsyes+\frac{\Delta}{4}$, 
a contradiction to $\criteps \leq \epsyes$.
So, the polygon computed in this call is a solution.

An overall complexity analysis of approximating $\criteps$ (and computing a solution)
is relatively straightforward: 
$I$ is obviously halved in every iteration, so it takes $O(\log(\frac{r}{\Delta}))$ iterations
to approximate $\criteps$.
Every iteration is bounded by the complexity given in Theorem~\ref{thm:rational_complexity}.
We omit further details of the proof:
\begin{theorem}\label{thm:criteps_complexity}
Approximating $\criteps$ to a precision $\Delta>0$ requires
$$O(n\frac{r}{\Delta}\sqrt{\frac{\criteps}{\Delta}}\cdot \log(n\frac{r}{\Delta}\sqrt{\frac{\criteps}{\Delta}}))$$
arithmetic operations with rational numbers.
\end{theorem}

\ignore{
\begin{figure}[h]
\centering
\subfloat[Exact results]{\label{fig:decide_exact}\includegraphics[width=0.4\textwidth]{decision_exact.eps}}
\subfloat[Rational results]{\label{fig:decide_rat}\includegraphics[width=0.4\textwidth]{decision_rational.eps}}
\caption{Decision procedure results versus monotone increasing $F = \criteps(r)$.}
\label{fig:decide_crit_eps}
\end{figure}
}

\myparagraph{Searching valid radii}
We assume now that $Q$ and $\eps$ are given, 
and discuss the question of what is the set $R$ of radii such that the deconstruction problem
has a solution. A priori, it is not clear what is the shape of $R$, but we will prove that it
is an interval of the form $[0,r^*]$. Having established this, we can apply another variant
of binary search to approximate the extremal value $r^*$.

In order to prove that $R$ is an interval, we prove first that the deconstruction problem can always be solved
for $Q$, $r$ and $\criteps$. In other words, we can replace the infimum in Definition~\ref{def:criteps}
by a minimum. The proof relies on two properties of infinite intersections of insets and offsets
that we show first.
\begin{lemma}\label{lemma:inset_infinite}
Let $(A_i)_{i\in\mathbb{N}}$ be a sequence of closed sets in $\R^2$. Then
$$\inset(\bigcap_{i=0}^{\infty}A_i,r)=\bigcap_{i=0}^{\infty}\inset(A_i,r).$$
\end{lemma}
\begin{proof}
The fact follows readily from the definition of insets: 
If $a\in \inset(\bigcap_{i=0}^{\infty}A_i,r)$, 
then $D_r(a)$ is contained in $\bigcap_{i=0}^{\infty}A_i$. In particular,
it is contained in $A_i$ for every $i$ which proves one inclusion. The other
direction is similar.
\end{proof}
\begin{lemma}\label{lemma:offset_infinite}
Let $(A_i)_{i\in\mathbb{N}}$ be a sequence of closed sets in $\R^2$
with $A_0\supseteq A_{1}\supseteq\ldots$. Moreover, let
$(\lambda_i)_{i\in\mathbb{N}}$ be a monotonously decreasing sequence of real numbers 
that converges to $\lambda\in\R$. Then
$$\offset(\bigcap_{i=0}^{\infty}A_i,\lambda)=\bigcap_{i=0}^{\infty}\offset(A_i,\lambda_i).$$
\end{lemma}
\begin{proof}
The ``$\subseteq$'' inclusion is straight forward, so we concentrate on the ``$\supseteq$'' part.
Fix some $b\in\bigcap\offset(A_i,\lambda_i)$. For every $i\in\mathbb{N}$, there
exists some $a_i\in A_i$ such that $(b-a_i)\in D_{\lambda_i}$. Now, the sequence $(b-a_i)_{i\in\mathbb{N}}$
is a bounded sequence in $\R^2$ (bounded by $D_{\lambda_0}$) and therefore has a convergent sub-sequence
by the well-known Bolzano-Weierstrass Theorem. Let $r$ denote the limit point of this subsequence.
In particular, the corresponding subsequence of $(a_i)_{i\in\mathbb{N}}$ converges to $a:=b-r$.
We show that $a\in\bigcap A_i$ and $r\in D_{\lambda}$ which suffices to prove the claim.

Assume that $a\notin\bigcap A_i$. 
Then, there is some $n_0$ such that $a\notin A_{n_0}$.
Since $ A_{n_0}$ is closed,
$d(a, A_{n_0})=:\eps>0$,
where $d$ is the Euclidean distance function.
Moreover, because each $A_n$ with $n\geq n_0$ is included in $A_{n_0}$,
$d(a,A_n)\geq \eps$.
Because $(a_i)_{i\in\mathbb{N}}$ converges to $a$, we can find some $N\geq n_0$ such that
$d(a,a_N)<\eps$.
However, $a_N\in A_N$, so 
$$d(a,A_N)\leq d(a,a_N)<\eps=d(a,A_{n_0})\leq d(a,A_N),$$
which is a contradiction. The fact that $r\in D_{\lambda}$ follows by a similar argument.
\end{proof}

\begin{theorem}\label{thm:criteps_has_a_solution}
For arbitrary $Q$ and $r$, and $\criteps$ as from Definition~\ref{def:criteps}, 
there exists a solution to the deconstruction problem.
\end{theorem}

\begin{proof}
Because of Corollary~\ref{cor:algorithm_correct}, 
we need to prove that
$$Q\subseteq Q'_{\criteps}:=\offset(\inset(\offset(Q,\criteps),r),r+\criteps).$$
Let $(\eps_i)_{i\in\mathbb{N}}$ be a monotone decreasing sequence of real numbers
that converges to $\criteps$. 
Because $\eps_i>\criteps$ for each $i$, \algdecide\ return \YES\ for each $\eps_i$, which is equivalent to
$$Q\subseteq Q'_{i}:=\offset(\inset(\offset(Q,\eps_i),r),r+\eps_i).$$
It is therefore sufficient to prove that 
$$Q'_{\criteps}=\bigcap_{i=0}^{\infty}Q'_i.$$
For that, we apply Lemma~\ref{lemma:offset_infinite}
on the constant sequence $(Q)_{i\in\mathbb{N}}$ and on $(\eps_i)_{i\in\mathbb{N}}$
to obtain
$$\offset(Q,\criteps)=\bigcap_{i=0}^{\infty}\offset(Q,\eps_i).$$
Applying Lemma~\ref{lemma:inset_infinite} yields
$$\inset(\offset(Q,\criteps),r)=\bigcap_{i=0}^{\infty}(\inset(\offset(Q,\eps_i),r),$$
and applying Lemma~\ref{lemma:offset_infinite} 
for the sequences $(\inset(\offset(Q,\eps_i),r))_{i\in\mathbb{N}}$
and $(r+\eps_i)_{i\in\mathbb{N}}$ yields
$$\underbrace{\offset(\inset(\offset(Q,\criteps),r),r+\criteps)}_{=Q'_{\criteps}} 
=   \bigcap_{i=0}^{\infty}\underbrace{\offset(\inset(\offset(Q,\eps_i),r),r+\eps_i)}_{=Q'_i}.$$
\end{proof}

Since $r$ is no longer fixed, we now consider $\criteps$ as a function 
from $\R^+$ to $\R^+$ depending on $r$; abusing notation, we will let 
$\criteps$ by itself denote this function from now on.

\begin{theorem}\label{thm:monotone_crit_eps}
$\criteps$ is a monotone increasing function.
Moreover, $\criteps$ is Lipschitz-continuous
with Lipschitz factor $1$.
\end{theorem}
%
\begin{proof}
We prove monotonicity first:
Let $a<b$ denote two radii and $\eps_{a}=\criteps(a)$, $\eps_{b}=\criteps(b)$. We will show that 
\algdecide$(a,\eps_{b})$ returns \YES, which proves that 
$\eps_{a}\leq\eps_{b}$.

We compare the intermediate results of \algdecide$(a, \eps_{b})$, denoted by $S_i$, 
to those of \algdecide$(b, \eps_{b})$, denoted by $B_i$:
\begin{flalign*}
S_1& =\offset(Q, \eps_{b})&
B_1& =\offset(Q, \eps_{b}) = S_1\\
S_2& =\inset(S_1, a)&
B_2& =\inset(B_1, b) = \inset(\inset(S_1, a) , b-a) = \inset(S_2, b-a)\\
S_3& =\offset(S_2, a+\eps_{b})&
B_3& =\offset(B_2, b+\eps_{b}) = \offset(\offset(B_2,b-a), a+\eps_{b})
\end{flalign*}
Since 
$$\offset(B_2,b-a) = \offset(\inset(S_2, b-a), b-a) \subseteq S_2$$
it follows that $B_3 \subseteq S_3$.
Because \algdecide$(b, \eps_{b})$ returns \YES\ by definition, 
it holds that $Q \subseteq B_3$, therefore $Q \subseteq S_3$
and \algdecide$(a, \eps_{b})$ also returns \YES.

For Lipschitz continuity, let $a<b$ be such that $b-a\leq\delta$, and again $\eps_{a}=\criteps(a)$, $\eps_{b}=\criteps(b)$.
We show that $\eps_{b}-\eps_{a}\leq \delta$.
There exists a
polygonal region $P$ that is a solution
to the deconstruction problem for $Q$, $a$ and $\eps_{a}$. In other words,
$$H(\offset(P,a),Q)\leq \eps_{a}.$$
Because of the general fact
$$H(A,B)\leq \eps \Rightarrow H(\offset(A,\delta),B)\leq \eps+\delta,$$
and since $b\leq a+\delta$ we have that
$$H(\offset(P,b),Q)  \leq  H(\offset(P,a+\delta),Q) 
                     =   H(\offset(\offset(P,a),\delta),Q) 
                   \leq  \eps_{a}+\delta.$$
Therefore, $P$ is a solution for the deconstruction problem for $Q$, $b$ and $\eps_{a}+\delta$,
so $\eps_{b}\leq\eps_{a}+\delta$.
\end{proof}

It follows from the monotonicity and Theorem~\ref{thm:criteps_has_a_solution} 
that $R$ is an interval which has $0$ as its left endpoint. 
Thus, computing $R$ reduces to
finding the maximal $r^*>0$ such that $\criteps(r^*)=\eps$.

Using the exact decision procedure, we can perform a binary search
similar to that for approximating $\criteps$: First, we compute
an interval $[0,r]$ containing $r^*$. Since $Q$ is finite, we can 
take $r$ to be the radius of the smallest enclosing circle of $Q$ plus $\eps$.
Then, we start the iterative process, deciding on the left or right subinterval
depending on the result of \algdecide\ for $Q$, $\eps$ 
and the midpoint of the interval.

What if we are using \algdecideapp instead?
Unlike Algorithm~\ref{alg:search_criteps}, we can no longer guarantee that
every execution of the approximation algorithm halves the search interval,
because a return value \UNDECIDED\ does not bound the distance 
of the current radius $r$ to the critical value $r^*$.
Instead, we propose the following scheme: 
For an interval $I$ with midpoint $r$,
\algdecideapp is called with some $\delta$, initially set to $\frac{\eps}{2}$.
If it returns \UNDECIDED, $\delta$ is divided by $2$ and \algdecideapp is recalled.
Eventually, the algorithm returns \YES\ or \NO, and the interval $I$ can be halved.

Let $R'$ be the preimage of $\eps$ under $\criteps$.
Note that $R'$ is an interval (which may consist of only one point).
The algorithm from above is guaranteed to converge to some $r\in R'$. 
However, if $R'$ contains more than one point,
it is not guaranteed to converge to the maximal one (because it
gets stuck in an infinite loop as soon as the query value $r$ lies in $R'$).
One way of avoiding this infinite loops is to decrease $\delta$
only to some threshold and choosing another query value $r$ from the interval
if no decision was made. 
Nevertheless, we have not found an algorithm 
with the formal guarantee of converging
to the largest value in $R'$ eventually.

\myparagraph{Searching for both $r$ and $\eps$}
\def\WFPolygon{{%
\includegraphics[width=0.46\textwidth]{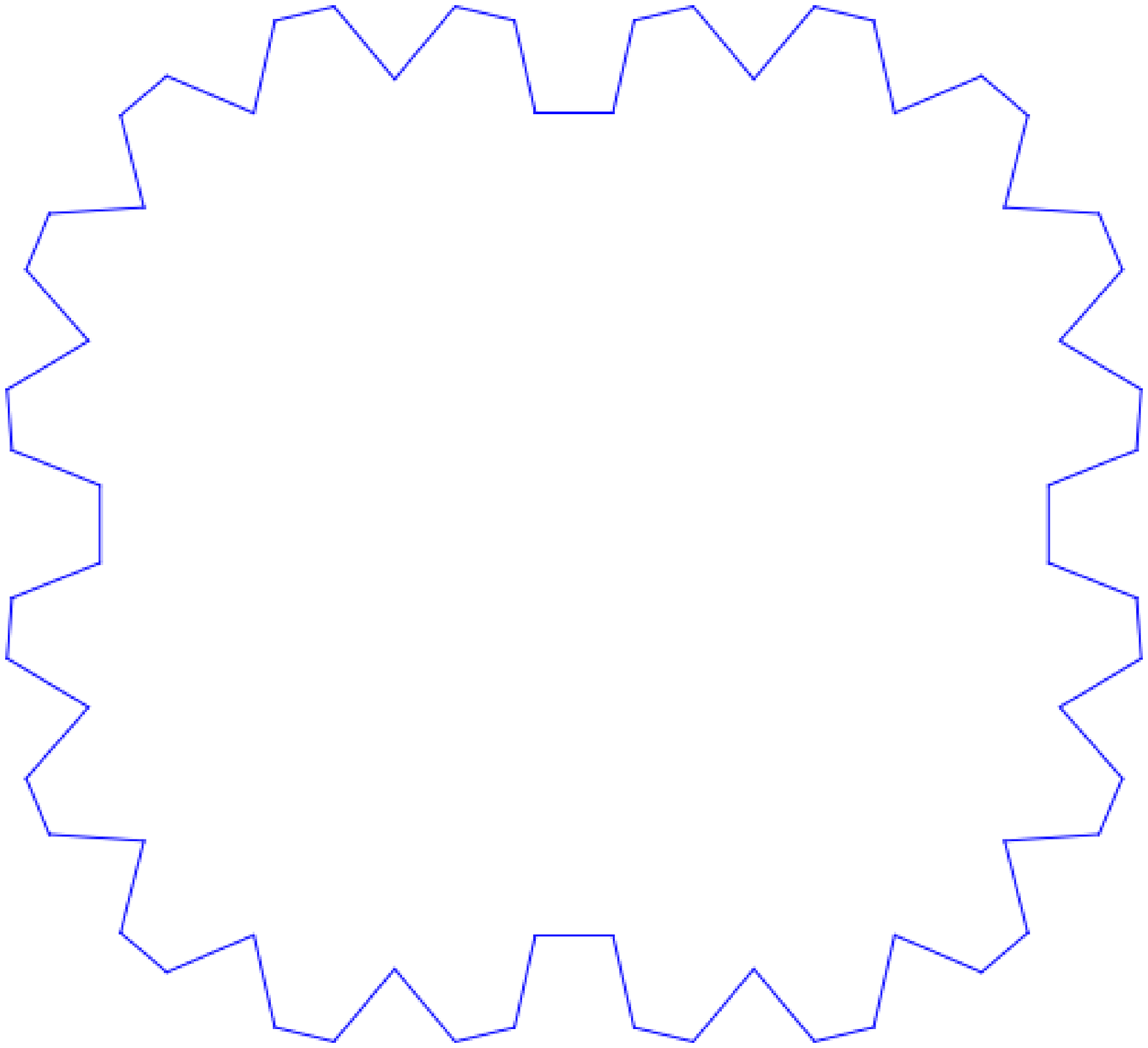}
}}
\ignore{
\def\WFCriteps{{%
\includegraphics[width=0.46\textwidth]{wf_critepsgraph.eps}
}}
\subfloat[The $\criteps(r)$ estimation with $\delta = \frac{1}{512}$]{
\label{fig:wf_criteps}
\WFCriteps
}}
\def\WFJGraph{{%
\includegraphics[width=0.46\textwidth]{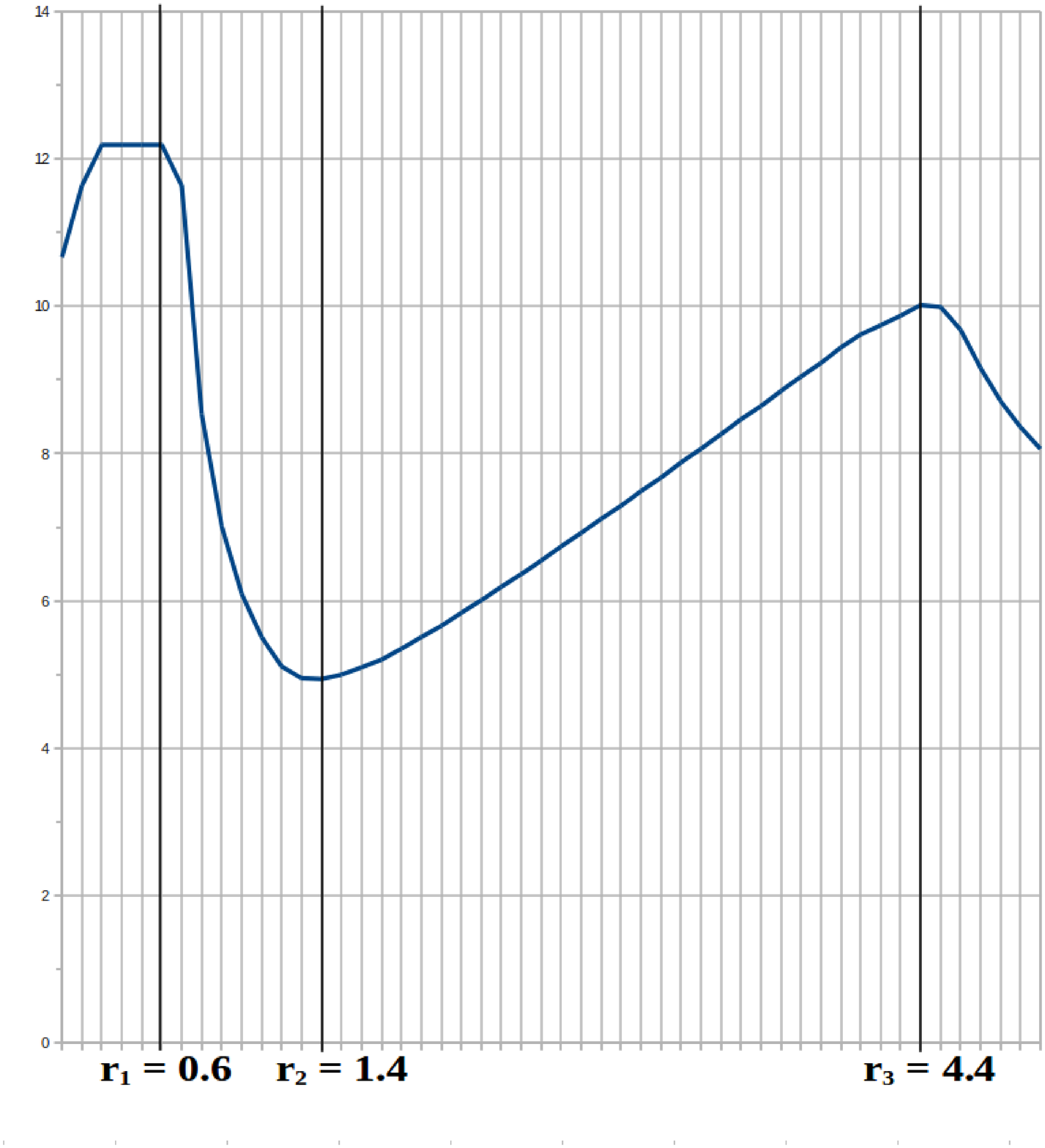}
}}
\begin{figure*}[p]
\subfloat[The polygon]{
\label{fig:wf_polygon}
\WFPolygon
}
\subfloat[The $J$-graph approximation with $\Delta=\frac{1}{512}$]{
\label{fig:wf_jgraph}
\WFJGraph
}
\caption{The Flower polygon example: $10$ samples per radius unit for $r\leq5$.
\label{fig:wf_example}
}
\end{figure*}
We finally consider the question of how we can find a reasonable choice of
$r$, $\eps$ and a polygonal region $P$, such that $P$ is a solution for 
the deconstruction problem for $Q$, $r$, and $\eps$.
The meaning of ``reasonable'' depends on the application context,
and possible prior information (for instance, a range of possible
offset radii). We offer a basic generic approach and justify our choice
with an example.

\def\WFMaxOneR{{%
\includegraphics[width=0.30\textwidth]{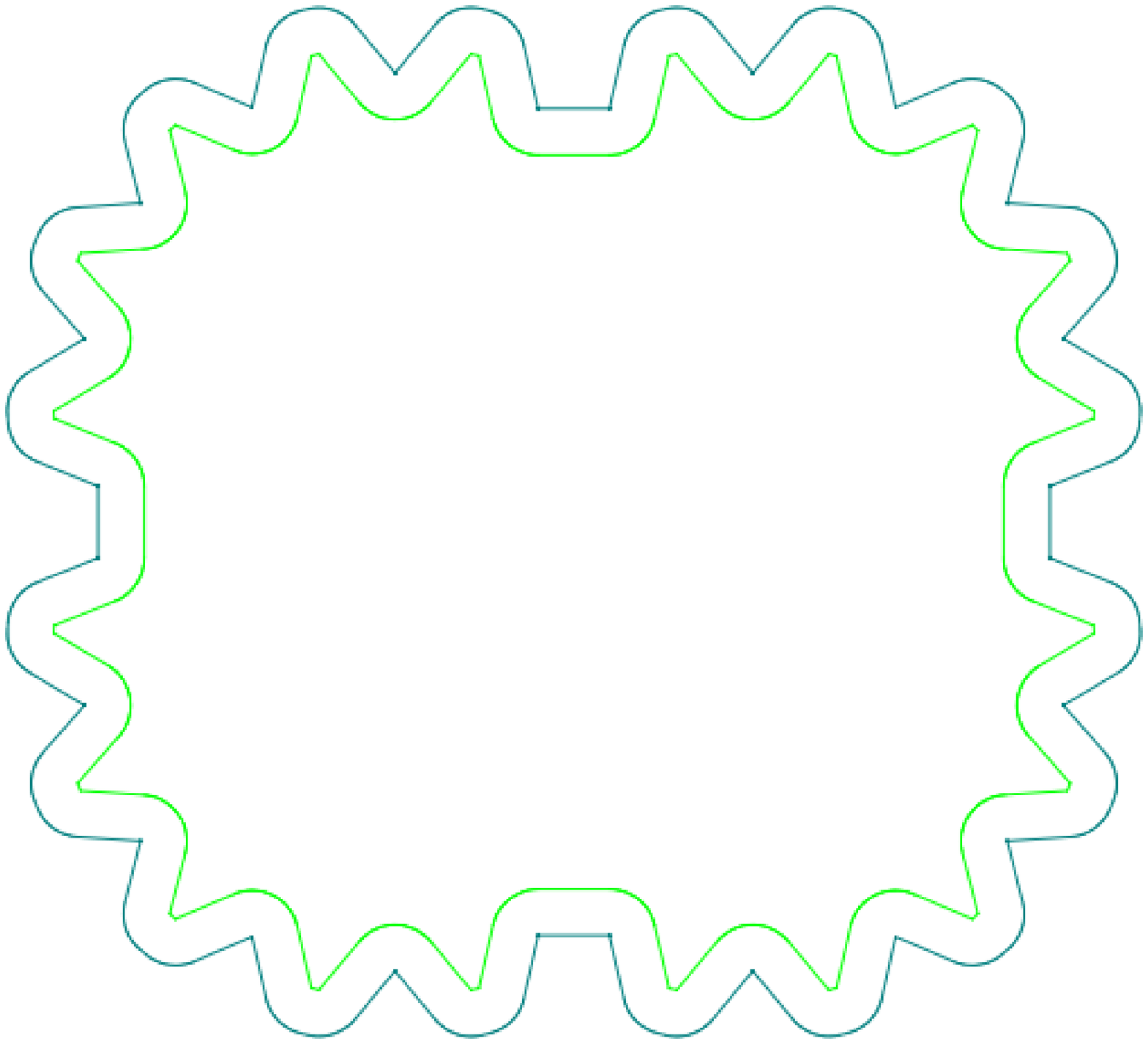}
}}
\def\WFMaxOneEps{{%
\includegraphics[width=0.30\textwidth]{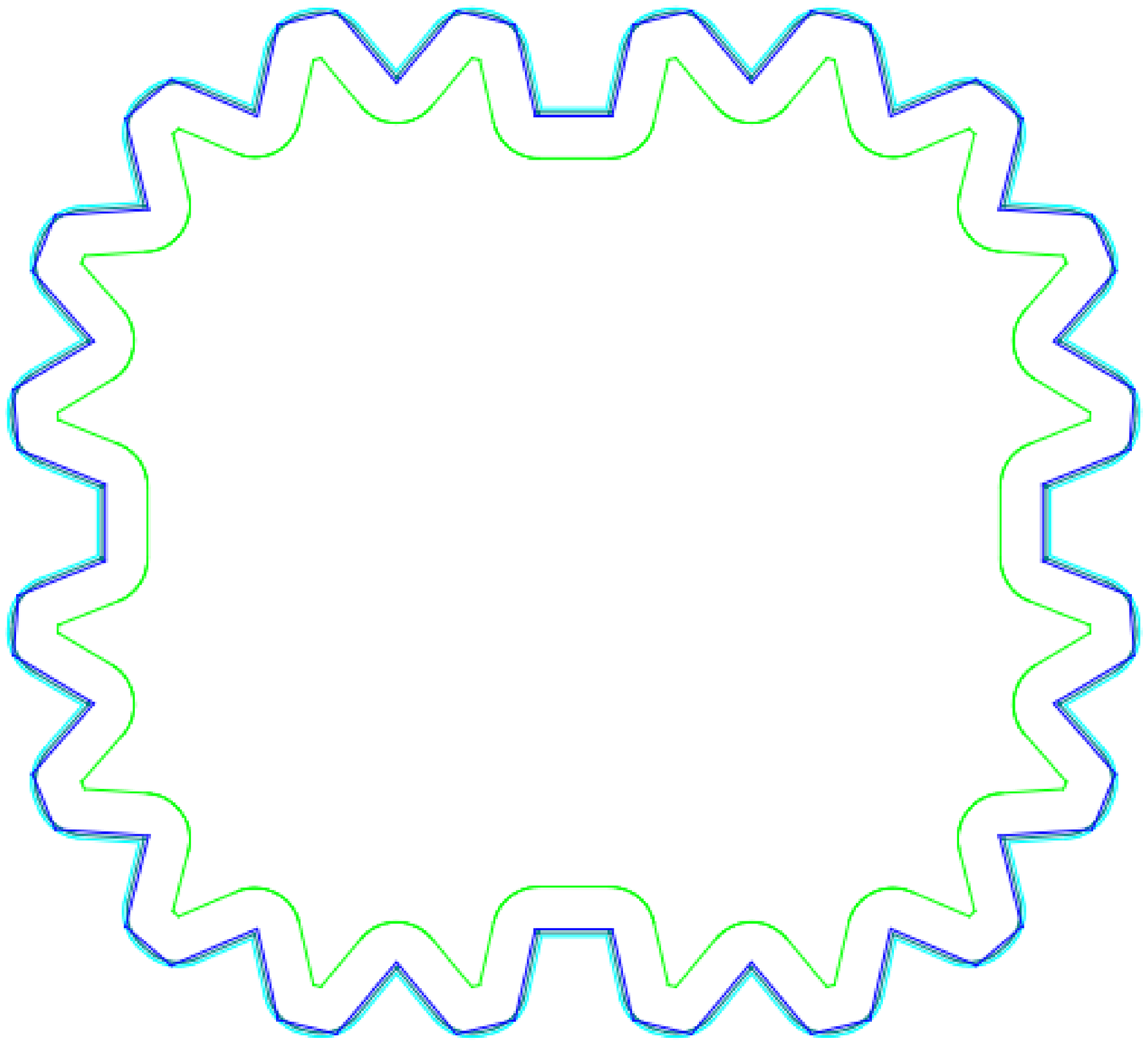}
}}
\def\WFMaxOneZoom{{%
\includegraphics[width=0.30\textwidth]{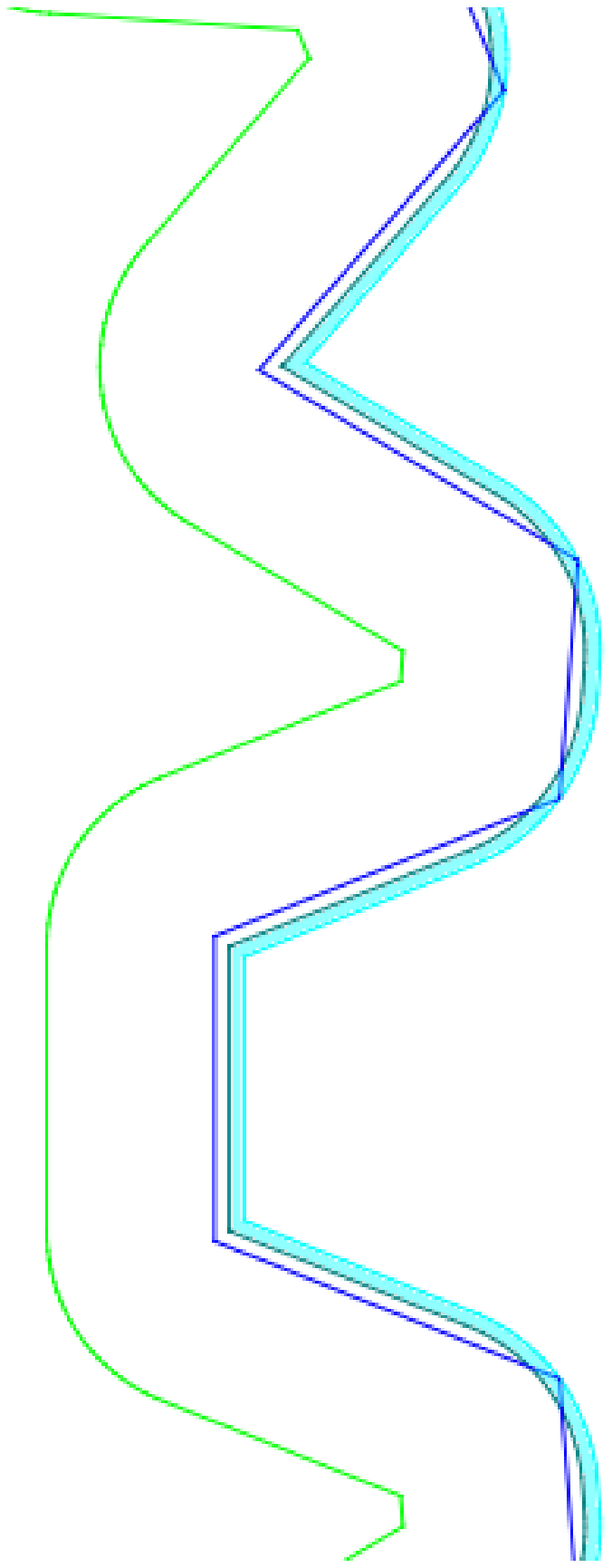}
}}
\def\WFMinR{{%
\includegraphics[width=0.30\textwidth]{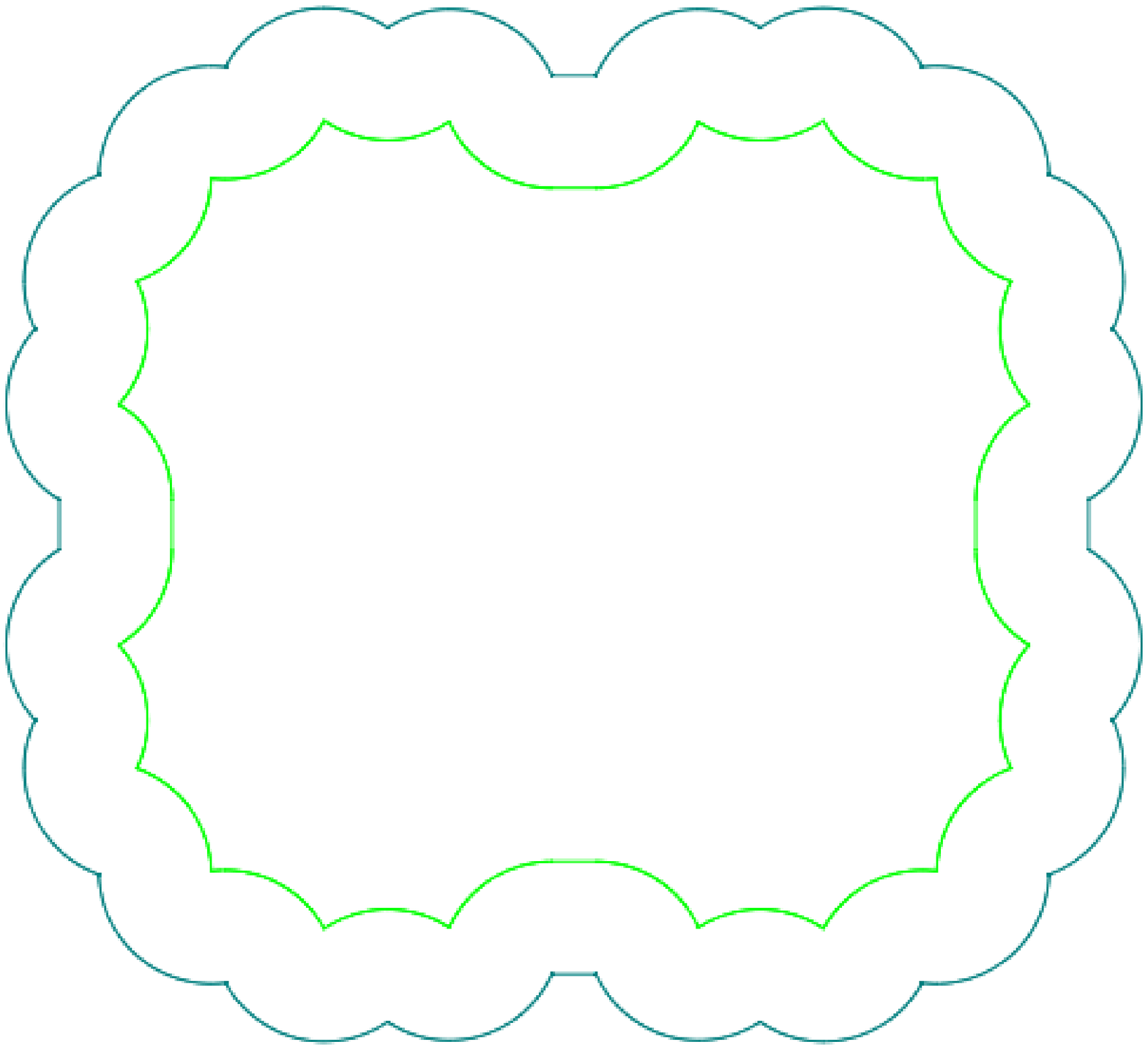}
}}
\def\WFMinEps{{%
\includegraphics[width=0.30\textwidth]{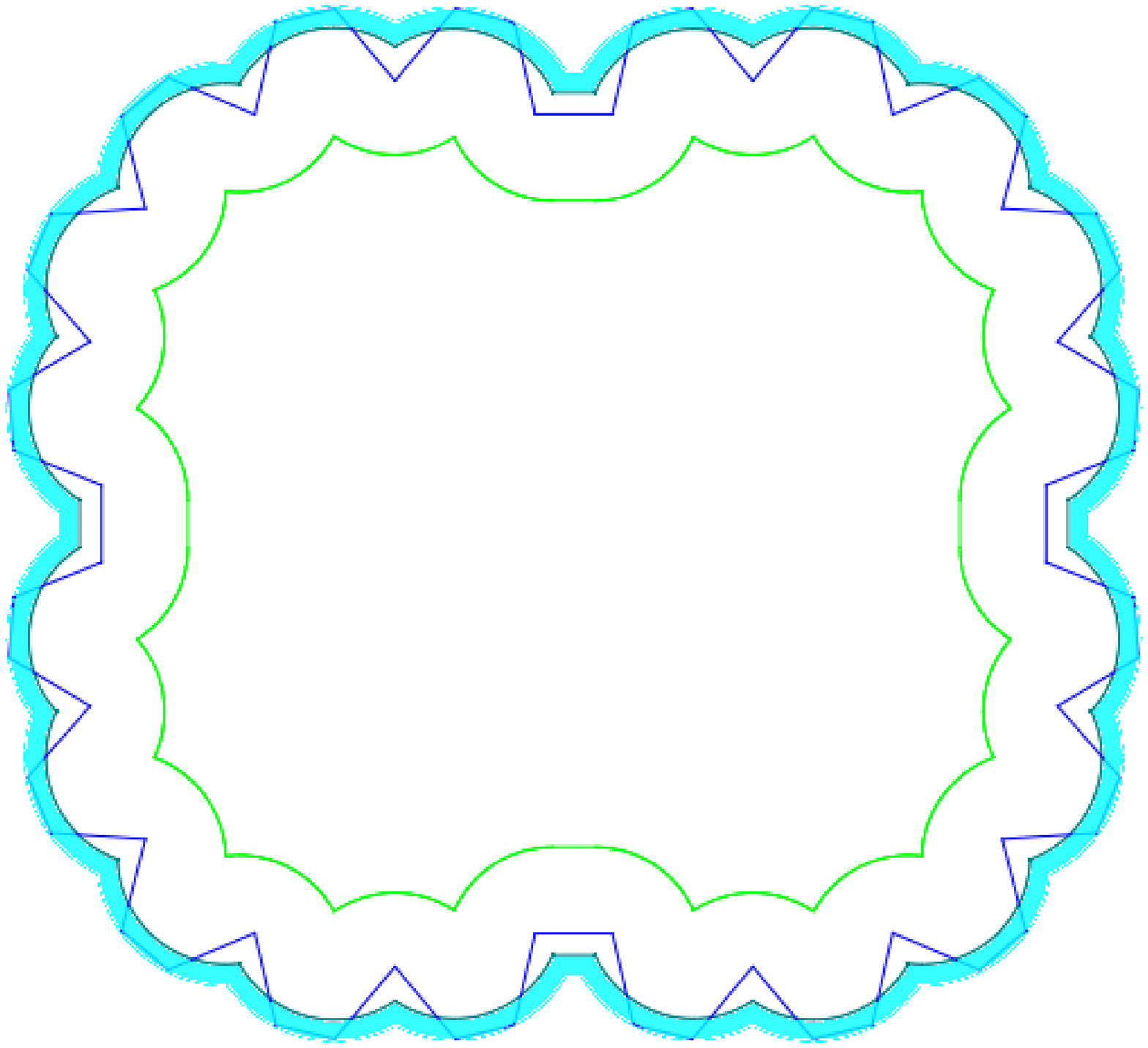}
}}
\def\WFMaxTwoR{{%
\includegraphics[width=0.30\textwidth]{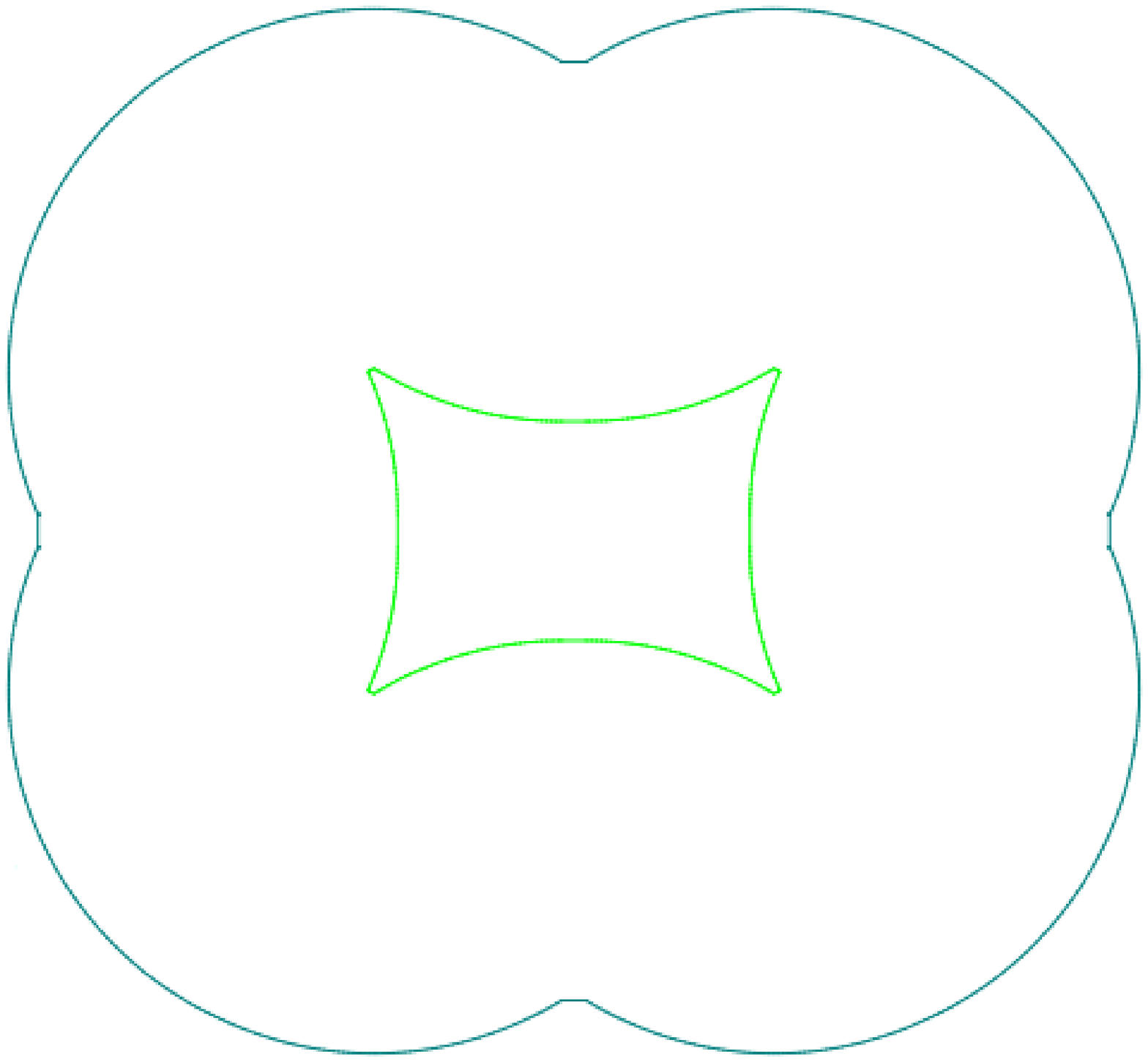}
}}
\def\WFMaxTwoEps{{%
\includegraphics[width=0.30\textwidth]{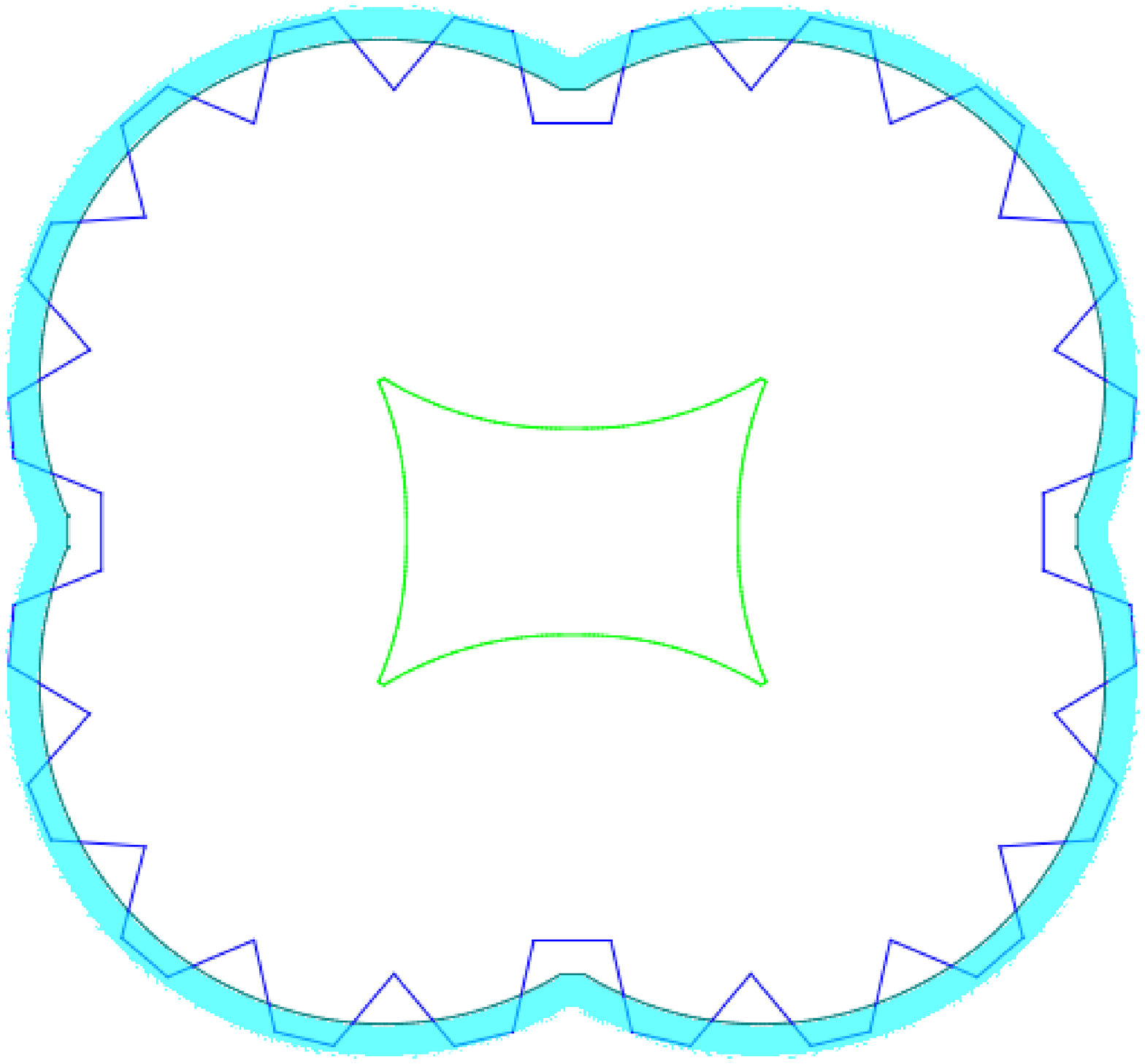}
}}
\begin{figure*}[p]
\subfloat[$J$ maximum at $r_1=0.6$]{
\label{fig:wf_max1_r}
\WFMaxOneR
}
\subfloat[$J$ minimum at $r_2=1.4$]{
\label{fig:wf_min_r}
\WFMinR
}
\subfloat[$J$ maximum at $r_3=4.4$]{
\label{fig:wf_max2_r}
\WFMaxTwoR
}

\ignore{$\critepsappd(r_1) \approx \frac{21}{256} \cdot r_1 = \frac{63}{1280}$}
\subfloat[$\critepsappd(r_1) \approx 0.082 \cdot r_1 = 0.049\ldots$]{
\label{fig:wf_max1_reps}
\WFMaxOneEps
\begin{pspicture}(2.0, -1.6)(2,2)
\newpsstyle{NoShadow}{ shadow=false }
\PstLens[LensHandle=false, LensSize=1.0, LensMagnification=0.8, LensStyleGlass=NoShadow](0.8,0.7){\WFMaxOneZoom}
\end{pspicture}
}
\ignore{$\critepsappd(r_2) \approx \frac{830}{4096} \cdot r_2 = \frac{2905}{10240}$}
\subfloat[$\critepsappd(r_2) \approx 0.202 \cdot r_2 = 0.283\ldots$]{
\label{fig:wf_min_reps}
\WFMinEps
}
\ignore{$\critepsappd(r_3) \approx \frac{409}{4096} \cdot r_3 = \frac{4499}{10240}$}
\subfloat[$\critepsappd(r_3) \approx 0.099 \cdot r_3 = 0.439\ldots$]{
\label{fig:wf_max2_reps}
\WFMaxTwoEps
}
{
\caption{Flower polygon approximations for the $(r,\critepsappd(r))$ values at $J$-graph extrema.
In the upper row, approximations of the solutions are shown in green, 
and their $r$-offsets ($\critepsappd(r)$-close to the input) in dark blue. 
In the lower row the input polygon is shown in blue. The $\critepsappd$-width
cyan stripe around the $r$-offset demonstrates the quality of the approximation.
} 
\label{fig:wf_extremum}
}
\end{figure*}

Generally, we expect from a reasonable pair $(r,\eps)$
that $\eps$ is small. So, in order to judge whether a good solution exists
for radius $r$, we consider $\criteps(r)$. However, $\criteps(r)$ being 
small (or equivalently, $\frac{1}{\criteps(r)}$ being large) is not
a good criterion, because $\criteps$ is monotone increasing
according to Theorem~\ref{thm:monotone_crit_eps}, so $r=0$ would always be
the best solution.\footnote{Note that this is also formally correct, because
$P:=Q$ is the perfect solution for $r=0$.} In order to remove the bias
towards small radii, we scale the objective function and consider
$$J:\R^+\rightarrow\R^+, r\mapsto \frac{r}{\criteps(r)}.$$
Note that $J$ is well-defined on the positive axis and continuous.
Moreover, we can approximate the graph of $J$ in any finite interval
of $r$-values by choosing a sample of the interval and approximating
$\criteps$ at each sample value using Algorithm~\ref{alg:search_criteps}.

We demonstrate by an example that the local maxima of~$J$ yield 
radii that lead to good deconstruction results. Consider the polygonal region
defined in Figure~\ref{fig:wf_polygon}, 
and its (approximated) $J$-graph in Figure~\ref{fig:wf_jgraph}.
We can identify two local maxima~$r_1$ and~$r_3$; we have plotted the corresponding
solutions in Figure~\ref{fig:wf_extremum}. Indeed, we see that  
for the large radius $r_3$, we obtain a relatively simple solution whose
offset blurs away the spikes of $Q$. 
For the smaller local maximum at $r_1$, we obtain
a solution with more details such that 
the spikes can be approximated almost perfectly with the given radius.
In contrast, the shape at the local minimum $r_2$ combines the disadvantages of the two
discussed cases: the solution is similarly complicated
as the $r_1$-solution (it contains flattened versions of all the spikes),
but its approximation quality is not significantly better than
for the $r_3$ solution which achieves the same with a much larger radius.

\section{Deconstructing Convex Polygons}
\label{sec:convex}

Assume that the input $Q$ to
Algorithm~\ref{alg:decision_general} is a convex polygon.
We first improve the decision algorithm such that it runs
in linear time (Algorithm~\ref{alg:decision_convex}). Then
we look for a polygon $P$ with a minimal number of vertices
($\OPT$) such that $Q$ is $\eps$-close to $\offset(P,r)$.
We give a simple linear-time algorithm that
produces a polygon with at most $\OPT+1$ vertices.

\begin{lemma}\label{lem:convex-tilde-P-computation}
If $Q$ is a convex polygonal region, then $\TP$, as
computed by \algdecide\
(Algorithm~\ref{alg:decision_general}), is also a convex
polygon, and it can be computed in $O(n)$ time.
\end{lemma}

\begin{proof}
$Q$ is the intersection of the half-planes bounded by lines that support the
polygon edges. Observe that~$\TP$ can be directly constructed from $Q$
by shifting each such line by $r-\eps$ inside the polygon,
which shows that $\TP$ is convex.
%
%
 For the time complexity, we divide the shifted edges of $Q$ into
 those bounding $Q$ from above, and those bounding $Q$ from
 below (we assume w.l.o.g.\ that no edge is vertical).
 Consider the former edges; the lines supporting those edges
 have slopes that are monotonously decreasing when
 traversing the edges from left to right. We have to compute
 their lower envelope; for that, we dualize by mapping
 $y=mx+c$ to $(m,-c)$, which preserves above/below relations,
 and compute the upper hull of the dualized points.
 Since we already know the order of the points in their
 $x$-coordinate, this can be done in linear time using
 Graham's
scan~\cite{a-aeach-79,g-eadch-72}.
 The same holds for the edges bounding $Q$ from below, taking the
 upper envelope/lower hull.
\end{proof}

\algdecide\ first computes $\TP$ and checks whether
$Q\subseteq\offset(\TP,r+\eps)$. We replace the latter step
for convex polygons: Let $q_1,\ldots,q_n$ be the vertices
of $Q$ (in counterclockwise order) and define $K_i =
D_{r+\eps}(q_i)$, namely the disk of radius $r+\eps$
centered at $q_i$. We check whether all these disks
intersect $\TP$:

\begin{algorithm}[H]
\caption{\algconvexdecidefull
}
\begin{compactenum}[(1)]
\item $Q_\eps\gets\offset(Q,\eps)$
\item $\TP\gets\inset(Q_\eps,r)$
\item if $K_i\cap \TP\neq\emptyset$ for all $i=1,\ldots,n$, return \YES,\\
      otherwise return \NO
\end{compactenum}
\label{alg:decision_convex}
\end{algorithm}

\begin{lemma}
\label{lem:all_regions_intersected} \algconvexdecide\
agrees with \algdecide\ on convex input polygons $Q$ and
runs in $O(n)$ time.
\end{lemma}
\begin{proof}
For correctness, it suffices to prove that $\offset(\TP,r)$
is $\eps$-close to $Q$ if and only if each $K_i$ intersects $\TP$:
Indeed, if any $K_i$ does not intersect $\TP$, then $q_i$ has distance
more than $r+\eps$ to $\TP$, so $Q$ is not $\eps$-close to the offset.
Otherwise, if each disk $K_i$ intersects $\TP$,
$\offset(\TP,r+\eps)$ contains each vertex of $Q$.
Since it is a convex set (as the Minkowski sum of two convex sets),
it also covers each edge of $Q$. Thus, $Q\subseteq\offset(\TP,r+\eps)$,
which ensures that $Q$ is $\eps$-close to the offset
by Proposition~\ref{prop:closeness}.

For the complexity,
Lemma~\ref{lem:convex-tilde-P-computation} shows that the
computation of $\TP$ runs in linear time. We still have to
demonstrate that the last step of the algorithm (checking 
for non-empty intersections) also takes a linear
time. Let $e_1,\ldots,e_m$ be the edges of $\TP$ (with
$m<n$). To check for an intersection of $K_i$ with $\TP$,
we traverse the edges and check for an intersection,
returning \NO\ if no such edge is found. However, if such
an edge, say $e_j$ was found, we start the search for an
intersection of the next disk $K_{i+1}$ at $e_j$, again
traversing the edges in counterclockwise order. Using this
strategy, and noting that $K_1,\ldots,K_n$ are arranged in
counterclockwise order around $\TP$, it can be easily seen
that we iterate at most twice through the edges of
$\TP$.\end{proof}

\myparagraph{Reducing the number of vertices}
We assume that $\offset(\TP,r)$ is $\eps$-close to~$Q$.
We prefer a simple-looking approximation
of~$Q$, thus we seek a polygon $P\subseteq\TP$ whose offset is
$\eps$-close to~$Q$, but with fewer vertices than~$\TP$.
Any such $P$ intersects each of the bulged regions of
radius $r+\eps$: $\region_i:=K_i\cap\TP, i=1,\ldots,n$. We
call these bulged regions $\TP$'s \emph{\regterm s}. The
converse is also true: Any \emph{convex} polygon
$P\subseteq\TP$ that intersects all \regterm s
$\region_1,\ldots,\region_n$ has an $r$-offset that is
$\eps$-close to $Q$.

The following observation is a simple consequence
of Proposition~\ref{prop:closeness}:
\begin{proposition}\label{prop:closeness-inclusion}
If $\offset(P,r)$ is $\eps$-close to $Q$, and
$P\subseteq P'\subseteq \TP$, then $\offset(P',r)$ is $\eps$-close
to~$Q$.
\end{proposition}

We call a polygonal region $P$ \emph{(vertex-)minimal},
if its $r$-offset is $\eps$-close to $Q$, and there exists no other
such region with fewer vertices.
Necessarily, a minimal $P$ must be convex~--~otherwise, its convex hull
$\conv(P)$ has fewer vertices and it can be seen
by Proposition~\ref{prop:closeness-inclusion} that
$\offset(\conv(P),r)$ is also $\eps$-close to $Q$.
By the next lemma, we can restrict our search to polygons with vertices on
$\partial\TP$.

\begin{lemma}\label{lem:boundary}
There exists a minimal polygonal region $P\subseteq\TP$ the vertices of which
are all on $\partial\TP$.
\end{lemma}

\begin{wrapfigure}[6]{r}{3.2cm}
\vspace{-0.3cm}
\hspace*{-4mm}
\resizebox{3cm}{!}{\input{local_change.pstex_t}}
\end{wrapfigure}
\textsc{Proof.} We
pull each vertex $p_i \not\in \boundary\TP$ in the
direction of the ray emanating from $p_{i-1}$ towards $p_i$
until it intersects $\partial\TP$ in the point $p_i'$
(dragging $p_i$'s incident edges along with it); see the
enclosed illustration. For $P' =
(p_1,\ldots,p_{i-1},p_i',p_{i+1},\ldots,p_m)$: $P\subseteq
P'\subseteq \TP$, $\offset(P',r)$ is $\eps$-close to $Q$ by
Proposition~\ref{prop:closeness-inclusion}.\qed

We call a polygonal
region $P$ \emph{good}, if $P\subseteq\TP$, all
vertices of $P$ lie on $\partial\TP$, and $P$ intersects each \regterm\
$\region_1,\ldots,\region_n$. Note that any good $P$ is convex.

\begin{definition}
\label{def:intervals} For two points
$\tp,\tp'\in\partial\TP$, we denote by
$[\tp,\tp']\subset\partial\TP$ all points that are met
when travelling along $\partial\TP$ from $\tp$ to $\tp'$
in counterclockwise order. Likewise, we define half-open
and open intervals $[\tp,\tp')$,  $(\tp,\tp']$,
$(\tp,\tp')$.
\end{definition}

\begin{wrapfigure}[10]{r}{2.6cm}
\vspace{-0.4cm}
\resizebox{2.6cm}{!}
{\input{spot_and_eyelet_and_horizon.pstex_t }}
\label{fig:spot_and_eyelet_and_horizon}
\end{wrapfigure}
Let $\region_i=K_i\cap\TP$ be $q_i$'s \regterm\ as before.
Consider $\region_i \cap \partial\TP$. 
The portion of that
intersection set that is visible from $q_i$ (considering
$\TP$ as an obstacle) defines a (ccw-oriented) interval
$[v_i,w_i]\subset\partial\TP$. We call $v_i$ the
\emph{\regstart} of the \regterm\ $\region_i$. Finally, for
$\tp,\tp'\in\partial\TP$, we say that the segment
$\segment{\tp}{\tp'}$ is \emph{good}, if for all \regstart
s $v_i\in (\tp,\tp')$, $\segment{\tp}{\tp'}$ intersects the
corresponding \regterm\ $\region_i$. 

The figure above illustrates these definitions: The segment
$\segment{p}{p'}$ is good, whereas $\segment{p}{p''}$ is
not good, because $v_2\in (p,p'')$, but the segment does not
intersect~$\region_2$.

\begin{theorem}\label{thm:good_polygon_means_good_edges}
Let $P$ be a convex polygonal region with all its vertices on $\partial\TP$.
Then, $P$ is good if and only if all its bounding edges are good.
\end{theorem}
\begin{proof}
We first prove that if all the edges of $P$ are good, then $P$ is good.
It suffices to argue that it intersects all
\regterm s $\region_1,\ldots,\region_n$. Let
$p_1,\ldots,p_k$ be the vertices of $P$ in counterclockwise
order.
Any \regstart\ $v_i$ of an \regterm\ $\region_i$ either
corresponds to some vertex $p_\ell$ of~$P$, or lies
inside some interval $(p_\ell,p_{\ell+1})$. Since
$\segment{p_\ell}{p_{\ell+1}}$ is good, it
intersects~$\region_i$.

For the converse, assume that
$\segment{p_\ell}{p_{\ell+1}}$ is not good, which encloses
with the interval $(p_\ell,p_{\ell+1})$ a polygonal region
$R\subseteq\TP\setminus P$. Hence, there is a \regstart\
$v_i\in R$ such that $\segment{p_\ell}{p_{\ell+1}}$ does
not intersect the \regterm\ $\region_i$. It follows that
the entire $\region_i$ is inside~$R$ (see the
illustration above, 
considering $\segment{p}{p''}$ and
$\region_2$). Thus, $P\cap\region_i=\emptyset$, and so $P$
cannot be good.
\end{proof}

For $\tp\in\partial\TP$, we define its \emph{horizon}
$h_\tp\in\partial\TP$ as the maximal point in counterclockwise direction
such that that segment $\segment{\tp}{h_\tp}$ is good. 
Consider again the figure above: 
The segment $\segment{p}{h_p}$ is tangential to
$\region_2$, so if going any further than $h_p$ on
$\boundary\TP$ from $p$, the segment would miss $\region_2$
and thus become non-good.

\begin{lemma}\label{lem:vertex_in_interval}
\label{lem:vertex_in_region} Let $P$ be a good polygonal
region, and $\tp\in\partial\TP$. Then, $P$ has a vertex
$p\in(\tp,h_\tp]$.
\end{lemma}
\begin{proof}
Assume to the contrary that $P$ has no such vertex, and let
$p_1,\ldots,p_\ell$ be its vertices on $\partial\TP$.
Let $p_j$ be the vertex of $P$ such that
$\tp\in(p_j,p_{j+1})$. Then, also $h_\tp\in (p_j,p_{j+1})$,
because otherwise, $p_{j+1}\in (\tp,h_\tp]$. Since $P$ is good, the
segment $\segment{p_j}{p_{j+1}}$ is good, too. It is not
hard to see that, consequently, both $\segment{p_j}{\tp}$ and
$\segment{\tp}{p_{j+1}}$ are good. However, the latter
contradicts the maximality of the horizon~$h_\tp$.
\end{proof}

For an arbitrary initial vertex $s\in\partial\TP$, we finally specify
a polygonal region $P^s$ by iteratively defining its vertices. Set
$p_1:=s$. For any $j\geq 1$, if the segment $\segment{p_j}{s}$, which
would close~$P^s$, is good, stop. Otherwise, set $p_{j+1}:=h_{p_j}$.
Informally, we always jump to the next horizon until we can reach~$s$
again without missing any of the \regterm s. By construction, all
segments of $P^s$ are good, so $P^s$ itself is good.
The (almost-)optimality of this construction mainly follows
from Lemma~\ref{lem:vertex_in_interval}.

\begin{theorem}\label{thm:opt_plus_one}
Let $P$ be a minimal polygonal region for $Q$, having $\OPT$ vertices.
Then, for any $s\in\partial\TP$, $P^s$ has at most $\OPT+1$ vertices.
\end{theorem}
\begin{proof}
We first prove that $P^s$ has the minimal number of vertices among all good
polygonal regions that have $s$ as a vertex.
Let $s:=p_1,\ldots,p_m$ be the vertices
of $P^s$. There are $m-1$ segments of the form
$\segment{p_{\ell}}{h_{p_\ell}}$.
By Lemma~\ref{lem:vertex_in_region}, any good polygonal region
has a vertex inside each of the intervals $(p_\ell,h_{p_\ell}]$.
Together with the vertex at $s$, this yields
at least $m$ vertices, thus $P^s$ is indeed minimal among these polygonal
regions.

Next, consider any minimal polygonal region $P^{\star}$. We
can assume that all its vertices are on $\partial\TP$ by
Lemma~\ref{lem:boundary}. If $s$ is not a vertex of
$P^{\star}$, we add it to the vertex set and obtain a
polygonal region $P'$ with at most $\OPT+1$ vertices that
has $s$ as a vertex. $P^s$ has at most as many vertices as
$P'$, so $m\leq\OPT+1$.
\end{proof}

As each visit of an \regterm{} requires constant time, the construction
of a horizon is proportional to the number of visited \regterm s, and
there are only linearly many \regterm s. Thus, we can state:

\begin{theorem}\label{thm:opt_in_linear_time}
For an arbitrary initial vertex $s$, computing $P^s$
requires $O(n)$ time.
\end{theorem}
\begin{proof}
We prove that computing the horizon of a point $u$ takes a
number of operations proportional to the number of \regterm
s that are visited by the segment $\segment{u}{h_u}$. Let
us consider an arbitrary $u\in\partial\TP$. By
rotating appropriately, we can assume, without loss of
generality, that $u$ lies on a vertical edge of $\TP$ (or,
if $u$ is a vertex, that the next edge in counterclockwise
order is vertical), and that the edge is traversed
top-down. The horizon is determined by the slope of the
edge at $u$. Note that for each \regterm\
$\region_1,\ldots,\region_n$, there is an interval of
slopes $I_1^{(u)},\ldots,I_n^{(u)}$ such that the segment
from $u$ with slope $\lambda$ intersects $\region_i$ if and only
if $\lambda\in I_i^{(u)}$. Furthermore, each single $I_i^{(u)}$
can be computed with a constant number of arithmetic
operations. Assuming that the next \regterm\ to be travelled
from the current $p_i$ is $\region_j$, we can iteratively
compute the intersections $I_j\cap I_{j+1}\cap
I_{j+2},\ldots$ until $I_{j}\cap\ldots\cap I_{j+k}$ is
empty. In this case, we choose $\lambda_i :=
\max(I_{j}\cap\ldots\cap I_{j+k-1})$ as the slope for the
next segment, which must be $\segment{p_i}{h_{p_i}}$ since
it is good by construction, and any larger slope would
produce a non-good segment. Based on this property, it is
easy to show that computing~$P^s$ needs a number of
operations which is proportional to $n$, the number of
\regterm s.
\end{proof}

\section{Open Problems}
\label{sec:conclusion}

We have shown how to decide whether a given arbitrary
polygonal shape~$Q$ is composable as the \Mswithbreak{} of
another polygonal region and a disk of radius $r$, up to some
tolerance~$\eps$. Many related questions remain open.
\begin{inparaenum}[(i)]
\item Deconstruction of Minkowski sums seems more difficult
when both summands are more complicated than a disk; many
practical scenarios may raise this general deconstruction
problem.
\item It would be interesting to analyze the
deconstruction not only under the Hausdorff distance but
for other similarity measures, such as the Fr{\^e}chet or
the symmetric distance.
\item Can one remove the extra vertex when seeking an
optimal (vertex minimal) polygonal summand~$P$ in the
convex case.
\item Finding an optimal or near-optimal
polygonal summand in the non-convex case seems challenging.
\item As in polygonal simplification, we could also search
for the polygonal region with a given number of vertices
whose $r$-offset minimizes the (Hausdorff)
distance to the given shape.
\item
The offset-deconstruction problem can be reformulated in
higher dimensions. We consider especially the
three-dimensional case to be of practical relevance.
\end{inparaenum}

\section*{Acknowledgements} We thank Eyal Flato (Plataine~Ltd.) 
 for raising the offset deconstruction problem in connection 
 with wood cutting. We also thank Tim Bretl (UIUC) for suggesting 
 the digital-pen offset deconstruction problem.
 This work has been supported in part by the
Israel Science Foundation (grant no. 236/06), by the
German-Israeli Foundation (grant no. 969/07), by the
Hermann Minkowski--Minerva Center for Geometry at Tel Aviv
University, and by the EU Project under Contract No.~255827
(CGL---Computational Geometry Learning). 

\bibliography{bib}
\bibliographystyle{plain}

\end{document}